\documentclass[12pt]{article}
\usepackage{epsfig}
\usepackage{amsmath}
\newcommand{\Sc}{\scriptstyle}
\newcommand{\bea}{\begin{eqnarray}}
\newcommand{\eea}{\end{eqnarray}}
\newcommand{\bp}{\begin{pmatrix}}
\newcommand{\ep}{\end{pmatrix}}

\newcommand{\Slash}[1]{#1 \hspace{-.5em}/\hspace{.11em}} 
\newcommand{\fourint}[1]{\int\!\frac{d^4 #1}{(2\pi)^4}}

\newcommand{\dpart}[2]{\frac{\partial #1}{\partial #2}}

\newcommand{\cdt}{\!\cdot\!}

\begin{document}

\begin{flushright}
 ADP-02/70-T510 \\
 UNITU-THEP-09/2002\\
 hep-ph/0204178
\end{flushright}
\smallskip

\begin{center}
\begin{large}
{\bf Nucleon form factors from a covariant quark core:\\ limits
in their description}\\
\end{large}
\vspace{1cm}
{\bf 
 M.~Oettel$^1$$^\dagger$$^*$ and R.~Alkofer$^2$
\\}
\vspace{0.2cm}
{$^1$ Special Research Centre for the Subatomic Structure of Matter, \\
University of Adelaide, Adelaide SA 5005, Australia \\
$^2$ Institute for Theoretical Physics, T\"ubingen University, \\
Auf der Morgenstelle 14, D-72076 T\"ubingen, Germany \\
}
\end{center}
%
%
\begin{abstract}
\normalsize
\noindent
 In treating the relativistic 3--quark problem, a dressed--quark
 propagator parameterization is used which is compatible with recent lattice 
 data and pion observables. Furthermore 2--quark correlations
 are modeled as a series of quark loops in the scalar and axialvector
 channel. The resulting reduced Faddeev equations are solved for nucleon
 and delta. Nucleon electromagnetic form factors are calculated in a fully
 covariant and gauge--invariant scheme. Whereas the proton electric form factor
 $G_E$ and the nucleon magnetic moments are described correctly, the neutron
 electric form factor and
 the ratio $G_E/G_M$ for the proton appear to be quenched. The influence
 of vector mesons on the form factors is investigated which amounts  to
 a 25 \% modification of the electromagnetic proton radii within this
 framework.
\end{abstract}



\noindent
\rule{5cm}{.15mm}

\noindent
$^\dagger$Supported by a Feodor--Lynen fellowship of
the Alexander-von-Humboldt foundation and the Australian Research Council.\\
$^*$Address after April 30: MPI f\"ur Metallforschung,
Heisenbergstr.\ 1, 70569 Stuttgart, Germany

\newpage

\section{Introduction}

In tackling the covariant bound state problem in QCD, models based
on a combined Dyson--Schwinger (DS) and Bethe--Salpeter (BS) approach 
have found widespread application, for recent reviews see 
refs.~\cite{Alkofer:2000wg,Roberts:2000aa}.
This approach has been most successful in describing light pseudoscalar mesons
and their electromagnetic properties. Starting from 
a suitable model for the gluon and the gluon--quark vertex in the
infrared, and using this model consistently for the $q-\bar q$ scattering 
kernel in the meson BS equation, these mesons retain their character
as both $q-\bar q$ bound states and Goldstone bosons. 
Masses \cite{Alkofer:2002bp}, decay 
constants \cite{Maris:1997tm} and form factors \cite{Maris:2000sk} 
are found to be in excellent
agreement with experimental data.
In these studies the so--called rainbow--ladder approximation is used
which consists in retaining the bare quark--gluon vertex and a 
gluon propagator. The latter is modeled with an enhancement at intermediate 
momenta which provides enough strength to generate a dynamical quark mass.

Along these lines the nucleon's bound state amplitude can be obtained
by solving a relativistic Faddeev equation which needs as input
the full solution for the $q-q$ scattering kernel. It is known
that in the rainbow---ladder approximation this kernel exhibits diquark poles
\cite{Praschifka:1989fd,Burden:1997nh}, with scalar ($0^+$) diquarks 
($\approx 0.7 - 0.8$ GeV) and axialvector ($1^+$) diquarks
($\approx 0.9$ GeV) having the lowest masses. Other diquark correlations
have much larger masses. Although these poles might correspond to unphysical 
asymptotic states (and indeed disappear when going beyond rainbow--ladder
\cite{Bender:1996bb,Hellstern:1997nv,Bender:2002as}) 
they give us a hint that $0^+$--$1^+$
quark--quark correlations are expected to be dominant in the nucleon.
This argument receives support from recent lattice 
calculations~\cite{Wetzorke:2000ez} 
and also explains the $u$--$d$ valence quark 
asymmetry observed in deep inelastic scattering 
\cite{Close:1988br,Mineo:2002bg}.

The full relativistic Faddeev problem is highly involved and has been solved
so far only for a NJL model in lowest order where the $q-q$ interaction
is pointlike and therefore separable \cite{Ishii:1995bu}. 
If the $q-q$ scattering (or $t$) matrix is separable,
the Faddeev equations reduce to a quark--diquark BS equation which can be 
solved exactly. Inspired by this idea, the $t$ matrix has been modeled
in such a fashion in ref.~\cite{Oettel:2000jj}. Retaining free 
massive quarks and $0^+$--$1^+$ diquarks,
electromagnetic, strong and weak nucleon form factors 
have been calculated, in good
agreement with experiment (except for the magnetic form factors). 
It is noteworthy that especially
the neutron electric form factor is positive and different from zero
which is in contrast to the valence quark contributions in many non-- 
or semi--relativistic quark models. (Due to the approximate $SU(6)$
symmetry of these models, it is consistent with zero almost by construction.) 
Despite the positive results of the above mentioned  models, 
the assumptions of free massive constituent
quarks and diquarks is certainly too simplistic from a QCD point of view.
Another line of approach has been taken by 
ref.~\cite{Bloch:1999ke}. In this study, a well--constrained parametrization
of the quark propagator is used which was obtained by fitting it to 
a number of soft and spacelike meson observables \cite{Burden:1996ve}. 
It exhibits the basic feature 
of DS solutions: a mass function $M(p^2)$ which is of the order of 400 MeV
in the infrared and which evolves into the perturbative limit for
$p^2 \to \infty$. Furthermore this parametrization has no poles
thereby mimicking confinement via the absence of a Lehmann representation.
Scalar diquarks and the dominant nucleon Faddeev amplitude have also been 
modeled with entire functions (i.e. pole--free) 
and the electromagnetic form 
factors have been calculated. The results (fitted to $G_E$ of the proton)
show also a positive neutron $G_E$ and enhanced magnetic moments due
to the dressed quark propagator which also leads, by use of the Ward--Takahashi 
(WT) identity, to a dressed quark--photon vertex.  
A drawback of this study is the lack of manifest 
electromagnetic gauge invariance. In order to maintain it the calculation
of currents between bound states has to proceed by gauging (i.e. minimal
coupling of the photon) to an interaction kernel and sandwiching
the result between bound states which are {\em solutions} to bound state
integral equations with exactly the same kernel 
\cite{Kvinikhidze:1999xn,Oettel:2000gc}.

In both studies \cite{Oettel:2000jj,Bloch:1999ke} the parametrization
of the $q-q$ $t$ matrix bears no relation to the quark propagator and
thus to the dynamics which causes chiral symmetry breaking. There
remains the 
possibility that the good results especially for the
neutron $G_E$ are rather a result of a clever parametrization of
the $q-q$ correlations than they reflect the underlying physics.
Besides the bulk of contributions to observables coming from a quark core,
one would also expect corrections to these mainly coming from the pion cloud.
Their non--negligeable magnitude is apparent in recent lattice
extrapolations to small quark (or pion) masses 
\cite{Young:2001nc,Leinweber:1999ig}, also recent covariant studies 
\cite{Hecht:2002ej,Oettel:2002}  
confirm
that the nucleon mass shift due to pions is at least $-200$ MeV, 
thereby indicating the percentage level of pionic corrections to nucleon 
observables.  

Therefore, we will present an extension of the quark--diquark picture
which, besides covariance and gauge--invariance, aims to include
several constraints which are available through lattice and
other QCD--phenomenological studies. Thus the number of free parameters
will be confined, in fact to one, and we are in the position to explore 
the limits of a covariant nucleon quark core picture. We start
from the main assumption to neglect three--quark irreducible interactions
to arrive at solvable Faddeev equations. Evidence for this assumption 
is admittedly scarce,
only in the limit of static quark sources lattice data 
\cite{Alexandrou:2001ip} seem to confirm a picture where flux--tubes
between each pair of the three quarks minimize the free energy of the 
three--quark system. 
Proceeding from this assumption, we
employ the above mentioned efficacious quark propagator 
parametrization which captures the essentials of the infrared behaviour
of quarks within  QCD
to calculate separable $0^+$ and $1^+$ diquark correlations
by summing quark loop polarization diagrams (Sect.~\ref{dqs}).
These correlations are employed to solve the nucleon and delta Faddeev
equations (Sect.~\ref{bse}). Parameters are fixed by the  masses
of nucleon and delta, leaving only one free parameter which is essentially
the extension of the diquarks. Form factors are calculated in a manifestly
gauge invariance preserving scheme. Here, the construction
of the diquark correlations ensures that the photon properly
resolves the diquark.
The dependence of the form factors on the
diquark width is investigated (Sect.~\ref{ff}). Finally, in Sect.~\ref{concl},
we draw our conclusions.

Throughout this paper
we work in Euclidean metric ($g^{\mu\nu}=\delta^{\mu\nu}$,
$\{\gamma^\mu,\gamma^\nu\} = 2\delta^{\mu\nu}$, $\gamma^{\mu\dagger} =
\gamma^\mu$).

\section{Diquark correlations}
\label{dqs}

\subsection{The quark propagator}

\begin{figure}
 \begin{center}
   \epsfig{file=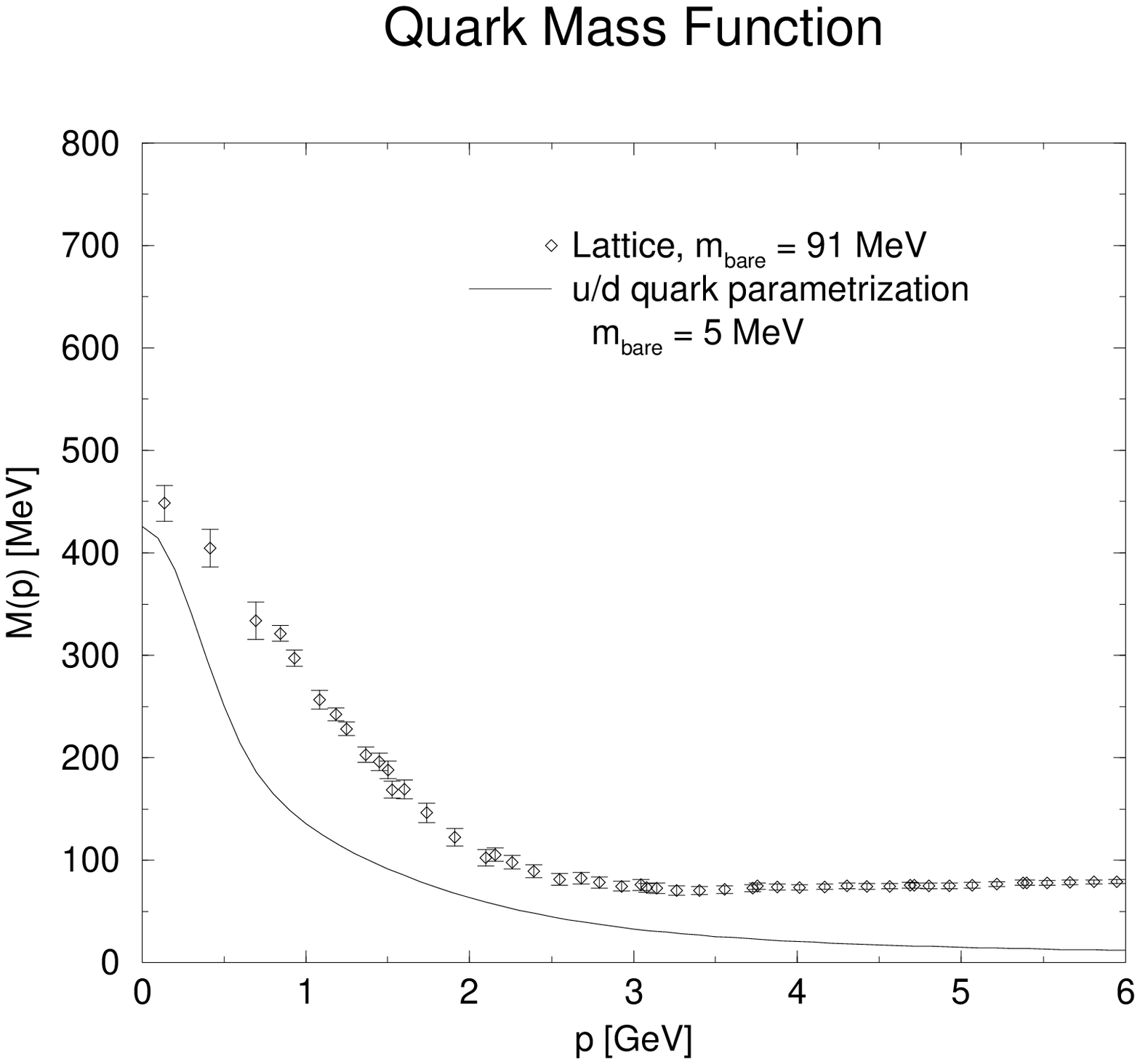,width=6cm}  \hskip 2cm
   \epsfig{file=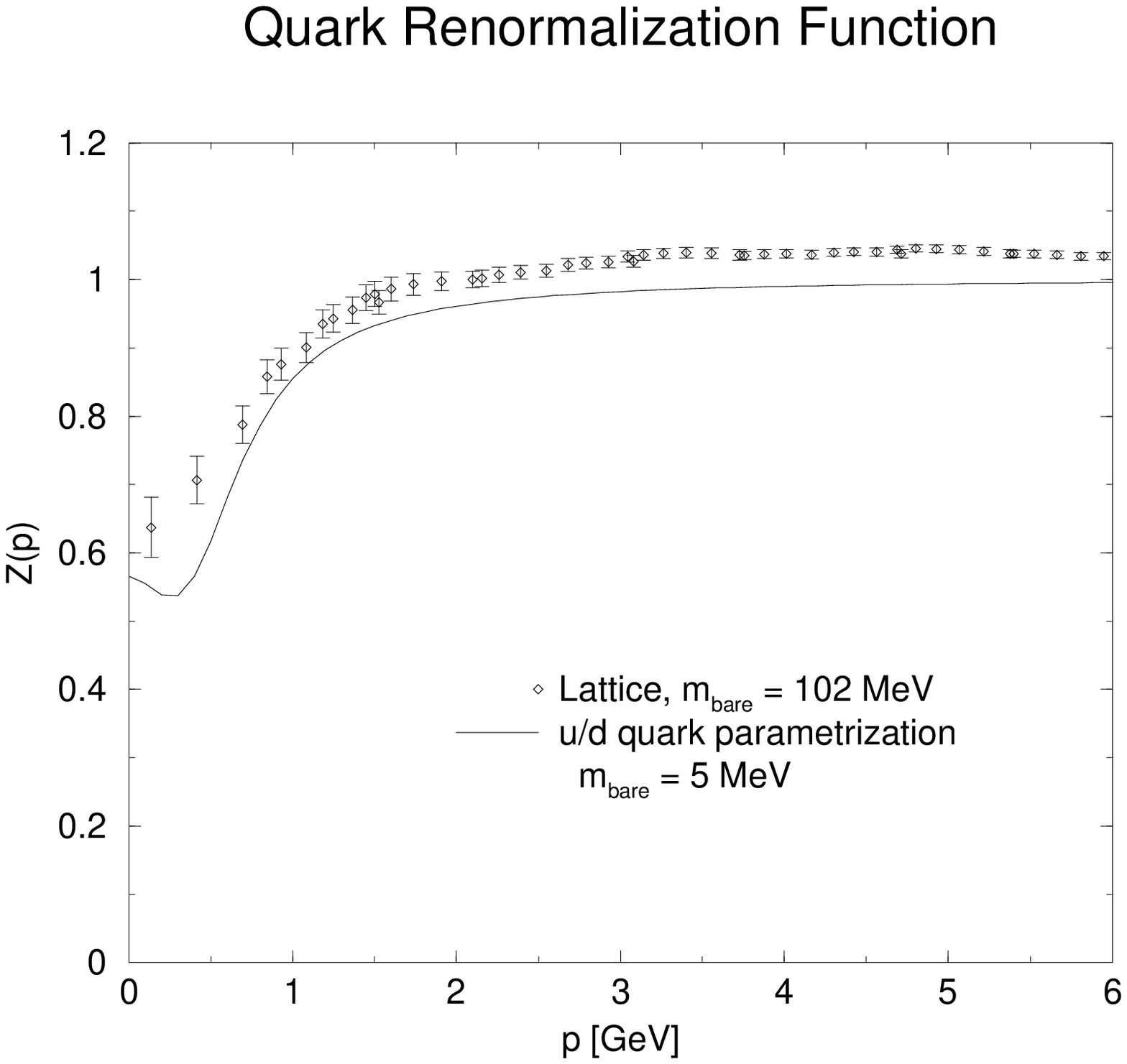,width=6cm} 
 \end{center}
 \caption{The quark mass function (left panel) and the renormalization function
 compared to lattice data.} 
 \label{qprop_f}
\end{figure}

Central ingredient to all calculations is the form of
the quark propagator, 
\begin{eqnarray}
\label{qprop}
S(p) & = & i\gamma\cdot p\, \sigma_V(p^2) - \sigma_S(p^2) \; ,\\
     & = & -\frac{Z(p^2)}{i\Slash{p}+M(p^2)} = -(i\Slash{p}A(p^2)+B(p^2))^{-1}
     \; . 
\end{eqnarray}
For the scalar and vector part we use the algebraic parametrizations:
\begin{eqnarray}
\label{ssm}
\bar\sigma_S(x) & =&  2\,\bar m \,{\cal F}(2 (x+\bar m^2))\\
&& \nonumber
+ {\cal F}(b_1 x) \,{\cal F}(b_3 x) \,
\left[b_0 + b_2 {\cal F}(\epsilon x)\right]\,,\\
\bar\sigma_V(x) & = & \frac{1}{x+\bar m^2}\,
\left[ 1 - {\cal F}(2 (x+\bar m^2))\right]\,,
\label{sigv}
\end{eqnarray}
with ${\cal F}(y) = (1-{\rm e}^{-y})/y$, $x=p^2/\lambda^2$, $\bar m$ =
$m/\lambda$,
$\bar\sigma_S(x)  =  \lambda\,\sigma_S(p^2)$   
and $\bar\sigma_V(x)  =  \lambda^2\,\sigma_V(p^2)$. 
The mass-scale is  $\lambda=0.566\,$GeV, and the parameter values
are given by
\begin{equation}
\label{tableA}
\begin{array}{ccccc} \hline \hline
   \bar m& b_0 & b_1 & b_2 & b_3 \\
  ~~~ 0.00897~~~ & ~~~0.131~~~ & ~~~2.90~~~ & ~~~0.603~~~  & ~~~0.185 ~~~\\ 
  \hline \hline 
\end{array}  ~~~.
\end{equation}
In Fig.~\ref{qprop_f} we show the quark mass function $M(p^2)$ and
the renormalization function $Z(p^2)$ for spacelike $p$ 
in comparison with recent lattice data that have been obtained in Landau 
gauge \cite{Skullerud:2001aw}.
Although the quark propagator fit has been performed to a number 
of meson observables
within the DS framework \cite{Burden:1996ve} and {\em not} to lattice data, 
the chosen parametrization represents the qualitative behavior
of both functions very well. It is still too premature to ask for quantitative
agreement since lattice calculations are not feasible for current quark masses
around 10 MeV yet. We note that the lattice data indicate
that the slope of the decreasing mass function is 
somewhat less steep than in the parametrization. We will find 
that this slope has influence on the ratio $G_E/G_M$ of the proton.
 
Both functions $\sigma_S$ and $\sigma_V$ are parametrized with entire 
functions. Thus they have no poles and reflect confinement. 
A major drawback, though, are the essential singularities at timelike 
infinity ($p^2=-\infty$). Consequently the quark renormalization function blows
up for timelike momenta, and it has been shown in ref.~\cite{Ahlig:2001qu} 
that this has disastrous consequences if one attempts to describe 
production processes where timelike momenta $O(1$ GeV$)$ are deposited 
onto the nucleon ($Z(-1 \mbox{GeV}^2)> 10^5$). Nevertheless,
for the bound state calculations described here, the quark propagator
is needed at complex momenta where always $|Z|<1$. Due to technical obstacles
we will calculate form factors only for $Q^2< 2$ GeV$^2$, and for these
calculations the quark propagator is sampled at momentum points where
$|Z|<1.2$. Thus for our calculations we do not expect artefacts  of the
parametrization to show up in the numerical results. 

\subsection{The $q-q$ $t$ matrix}

According to the arguments given in the Introduction, we expect scalar
and axialvector $q-q$ correlations to be the most important ones 
within the nucleon. Thus we model a separable $t$ matrix by
\begin{eqnarray}
 t(k_\alpha,k_\beta;p_\alpha,p_\beta) \equiv t(k,p,P) &=&
 \chi^5_{\alpha\beta}(k,P) \;D(P)\;\bar
  \chi^5_{\gamma\delta}(p,P) \; +\; \\
  & &
\chi_{\alpha\beta}^\mu(k,P) \;D^{\mu\nu}(P)\;
    \bar  \chi_{\gamma\delta}^\nu(p,P)  \; .
  \label{tsep}
\end{eqnarray}
The relative momenta are defined as
\begin{equation}
 \label{dq_rm}
 k[p]= \frac{1}{2}( k_\alpha[p_\alpha]-  k_\beta[p_\beta])\;,
\end{equation}
and the total diquark momentum is
\begin{equation}
 P=p_\alpha+p_\beta=k_\alpha+k_\beta\;.
\end{equation}
We assume that the Dirac structure of the vertices $\chi^5$ (scalar diquark)
and $\chi^\mu$ (axialvector diquark) is described by their 
leading components which non--relativistically correspond to quarks
being in a relative $s$ state:
\bea
  \label{chi5def}
  \chi^5_{\alpha\beta}(p)&=&g_{0^+} (\gamma^5 C)_{\alpha\beta}\; 
     {\cal F}(p^2/w_{0^+}), \\
  \label{chimudef}
  \chi^\mu_{\alpha\beta}(p)&=&g_{1^+} (\gamma^\mu C)_{\alpha\beta}\; 
    {\cal F}(p^2/w_{1^+}) \; .
\eea
The scalar function ${\cal F}$ (defined below eq.~(\ref{sigv})) 
describes the extension of the diquarks in
their relative momentum variable, regulated by the widths $w_{0^+}$
and $w_{1^+}$. Although the choice of this function is somewhat arbitrary,
numerical results depend only on the diquark widths and not on the specific 
form chosen as we have checked by employing both monopole and dipole forms.
The constants $g_{0^+}$ and $g_{1^+}$ are normalization constants yet
to be determined.

\begin{figure}
 \begin{center}
   \epsfig{file=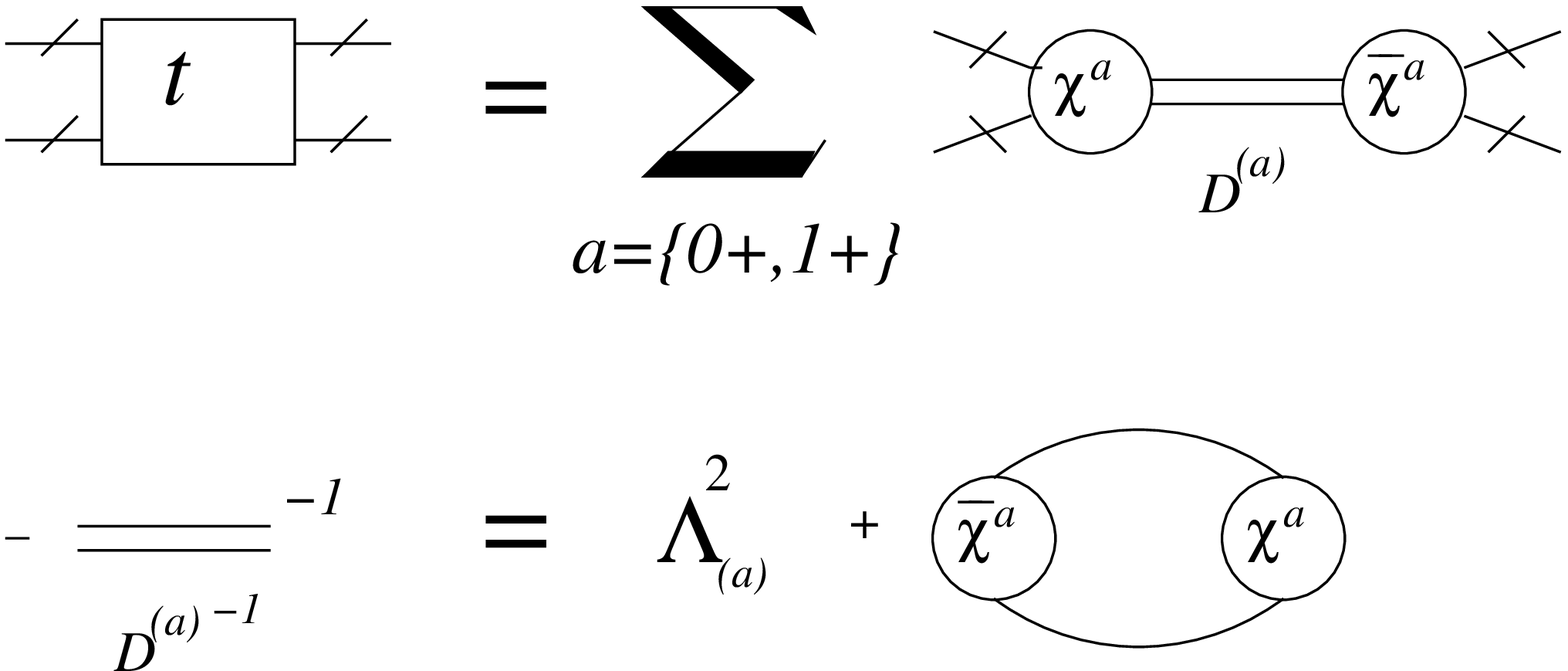,width=10cm} 
 \end{center}
 \caption{\label{tsep_f} 
 The separable $t$ matrix and the definition of the diquark 
  propagator}
\end{figure}

Antisymmetry between the quarks dictates the color and flavor
quantum numbers of the vertices $\chi$. They are both in a color antitriplet 
representation. As $(\gamma_5C)$ is antisymmetric, the scalar vertex must 
also be antisymmetric in flavor space, in contrast to the axialvector
vertex which is symmetric in flavor space due to the symmetric 
matrices $(\gamma^\mu C)$. We adopt the following normalizations,
restricting ourselves to the isospin subgroup:
\bea
 (\chi^{5,C})^{AB}_{ab} &=& \frac{(\tau_2)_{ab}}{\sqrt{2}}\;
 \frac{\epsilon^{ABC}}{\sqrt{2}} \; , \\
 (\chi^{\mu,C}_k)^{AB}_{ab} &=& \frac{(\tau_2\tau_k)_{ab}}{\sqrt{2}}\;
 \frac{\epsilon^{ABC}}{\sqrt{2}} \; .
\eea
Capital letters denote color indices and small letters isopin indices.
The $\tau^k$ represent the usual Pauli matrices. We will suppress
color and flavor indices in the following, corresponding traces will have
been worked out.

We model the inverse diquark propagators $D^{-1}$ and $(D^{\mu\nu})^{-1}$ by 
quark polarization diagrams as shown in Fig.~\ref{tsep_f} with an additional
constant offset $\Lambda_{0^+}$ for the scalar diquark and $\Lambda_{1^+}$ 
for the axialvector diquark. This assumption describes the propagation
of the quark pair being determined by an infinite series of loops as in
Fig.~\ref{tsep_f}. Similar expressions are obtained in the bosonized forms
of the Global Color Model \cite{Roberts:1988xc} 
or in the NJL model \cite{Ishii:1995bu}. In the latter
the constant $\Lambda$ corresponds to the inverse strength of the four--quark
interaction. 

For the scalar channel the inverse propagator reads
\bea
 D^{-1}(P^2) & = & - \Lambda_{0^+}^2 - \Pi(P^2)\; , \\
  \Pi(P^2) &=& -\fourint{q}{\rm Tr}\; \bar\chi^5(q^2)\;S(P/2+q)\;\chi^5(q^2)
   \;S^T(P/2-q)  \; .
\eea
The polarization function $\Pi(P^2)$ has also an essential singularity at
timelike infinity (as the quark propagator). Typically it evolves from
$-\infty$ at $P^2=-\infty$ monotonically to zero at $P^2=\infty$, save for
a tiny bump at around 4 GeV$^2$. The effect of the constant $\Lambda_{0^+}^2$
is to shift $D^{-1}$ downwards that it acquires a zero at some 
$-P^2=m_{0^+}^2$. Thus, the propagator has a pole. This is very similar 
to the rainbow--ladder truncation
of the quark DS eq./diquark BS eq. where poles do appear in the $t$ matrix.
At the pole we demand unit residue,
\begin{equation}
 \label{normsc}
   \frac{\rm d}{{\rm d}P^2}
     \Pi(P^2)|_{P^2=-m_{0^+}^2} \stackrel{!}{=} 1\;,
\end{equation}
thus relating the hitherto unknown constant $g_{0^+}$ to $m_{0^+}$
or, equivalently, $\Lambda_{0^+}$.

For the inverse propagator in the axialvector channel, we employ
an {\em ansatz} similar to the scalar channel:
\bea
 (D^{-1})^{\mu\nu}= -\Lambda_{\rm ax}^2\: \delta^{\mu\nu} - \Pi_{\rm ax}^{\mu\nu}\; .
\eea
The polarization loop
\bea
  \Pi_{\rm ax}^{\mu\nu}(P^2) &=& -\fourint{q}{\rm Tr}\; \bar\chi^\mu(q^2)\;
   S(P/2+q)\;\chi^\nu(q^2)
   \;S^T(P/2-q)  \; 
\eea
can be split into longitudinal ($P^\mu P^\nu/P^2\,\Pi_{{\rm ax,L}}(P^2)$) 
and transverse ($(\delta^{\mu\nu}-P^\mu P^\nu/P^2)\,\Pi_{{\rm ax,T}}(P^2)$)
components. We note that
$\Pi_{{\rm ax,L}}(0)=\Pi_{{\rm ax,T}}(0)$ as it should be (to have no pole
in the propagator at $P^2=0$). Furthermore we see from the numerical results
that in the region $P^2 \in [-0.7,1.5]$ GeV $\Pi_{{\rm ax,L}}(P)$ is
approximately constant (deviations are $\approx$ 1 \%). Therefore
we take the inverse propagator as
\bea
 (D^{-1})^{\mu\nu}(P)= -\Lambda_{\rm ax}^2 \delta^{\mu\nu} 
   - \Pi_{{\rm ax,T}}(P) \left(\delta^{\mu\nu}-\frac{P^\mu P^\nu}{P^2}\right) -
    \Pi_{{\rm ax,T}}(0)\,\frac{P^\mu P^\nu}{P^2}  \; .
\eea
This is in accordance with the requirement that  the longitudinal part of a
spin-1 propagator not be dressed.
 
The behavior of the transverse polarization, $\Pi_{{\rm ax,T}}$, is very 
similar to the scalar polarization, $\Pi$, thus upon shifting by 
$\Lambda_{\rm ax}^2$ the propagator acquires a pole at a mass which is 
larger than the scalar diquark mass if $\Lambda_{0^+}\sim\Lambda_{1^+}$.
At the pole a similar condition to eq.~(\ref{normsc}) holds:
\begin{equation}
 \label{normax}
   \frac{\rm d}{{\rm d}P^2}
     \Pi_{{\rm ax,T}}(P^2)|_{P^2=-m_{1^+}^2} \stackrel{!}{=} 1\;,
\end{equation}
We wish to relate the constants $\Lambda_{0^+}$ and $\Lambda_{1^+}$
which represent inverse coupling strengths in the scalar and axialvector 
channels, respectively.
Consider a quark vector current--current interaction
where the currents are color octet and Lorentz diagonal
which  arises e.g. from bosonizing
a Global Color Model with the gluon propagator in Feynman gauge,
$\sim \delta^{\mu\nu}$.
Upon a Fierz transformation into the scalar and axialvector diquark channels
we find  the relation \cite{Alkofer:1995mv}
\begin{equation}
 \label{NJL}
 \frac{\Lambda_{1^+}}{g_{1^+}}=2\;\frac{\Lambda_{0^+}}{g_{0^+}}\; .
\end{equation} 

In summary, we have parametrized the $q-q$ correlations close to the 
rainbow--ladder truncation scheme which proved to be successful in
the meson channels. Yet, scalar and axialvector diquarks are
mainly characterized by an (unphysical) mass which should only be interpreted
as an inverse effective correlation length than as a physical particle's
mass. The
composite nature of the diquarks is reflected by the propagators which are
a series of quark loops. We used six parameters (diquark widths $w_i$,
diquark normalization constants $g_i$ and
quark--quark inverse coupling strengths $\Lambda_i$, $i=\{0^+,1^+\}$) 
which are reduced to three free parameters by using the
relations (\ref{normsc},\ref{normax},\ref{NJL}). In the following,
we will impose two more constraints using the masses of nucleon and delta,
fixing essentially the inverse coupling strengths.

\section{Faddeev equations for nucleon and delta}
\label{bse}

A full derivation of the Faddeev equations for $N$ and $\Delta$ using
separable $q-q$ $t$ matrices  can be found
in ref.~\cite{Oettel:2000ig}. In the following, we will only 
introduce the necessary elements. For the case of separable $t$ matrices,
it is convenient to introduce baryon--quark--diquark Faddeev amplitudes.
These Faddeev amplitudes have to be decomposed in Dirac and Lorentz space
after their projection onto positive energy states with spin 1/2 ($N$)
or spin 3/2 ($\Delta$).

\subsection{Nucleon}

The nucleon Faddeev amplitude (or wave functions) can be described by
an effective multi-spinor characterizing the
scalar and axialvector correlations,
\begin{equation}
 \label{psidef}
 \Psi (p,P) u (P,s) \equiv
    \begin{pmatrix} \Psi^5 (p,P) \\ \Psi^\mu  (p,P) \end{pmatrix} u(P,s).
\end{equation}
$u(P,s)$ is a positive-energy Dirac spinor (of spin $s$), $p$ and $P$ are the
relative and total momenta of the quark-diquark pair, respectively.
The vertex functions are defined by truncation of the legs,
\begin{equation}
 \begin{pmatrix} \Phi^5  \\ \Phi^\mu \end{pmatrix} =
    S^{-1}  \begin{pmatrix} D^{-1} & 0 \\ 0 & (D^{\mu\nu})^{-1} \end{pmatrix}
 \begin{pmatrix} \Psi^5  \\ \Psi^\nu \end{pmatrix} .
 \label{amp}
\end{equation}
The coupled system of Faddeev equations for the nucleon wave and  vertex
functions can be written in the following compact form,
\begin{equation}
  \fourint{k} G^{-1}(p,k,P)
  \begin{pmatrix}\Psi^5 \\ \Psi^{\mu'}\end{pmatrix}(k,P) =0 \;,
  \label{bse_nuc}
\end{equation}
in which $G^{-1}(p,k,P)$ is the inverse of the full quark-diquark 4-point
function. It is the sum of the disconnected part and
the interaction kernel.

Here, the interaction kernel results from the reduction of
the Faddeev equation for separable 2-quark correlations.
It describes the exchange of the quark with one of those in the
diquark and thus the Faddeev equation reduces to an effective 
quark--diquark BS equation.
Thus,
\begin{eqnarray}
 G^{-1} (p,k,P) &=&
    (2\pi)^4 \;\delta^4(p-k)\; S^{-1}(p_q)\;
      \begin{pmatrix} D^{-1}\!(p_d)  & 0 \\ 0 & (D^{\mu'\mu})^{-1}\!(p_d)
         \end{pmatrix} - \nonumber\\
 & &\mbox{\hskip -1.8cm}  \frac{1}{2}
  \begin{pmatrix} - \chi^5{\Sc (p_2^2) } \; S^T{\Sc (q) }\; \bar\chi^5{\Sc
  (p_1^2) } &
     \sqrt{3}\; \chi^{\mu'}{\Sc (p_2^2) }\; S^T{\Sc (q) }\;\bar\chi^5 {\Sc
  (p_1^2) } \\
    \sqrt{3}\;\chi^5{\Sc (p_2^2) }\; S^T{\Sc (q) }\;\bar\chi^{\mu}{\Sc (p_1^2) }
    &  \chi^{\mu'}{\Sc (p_2^2) }\; S^T{\Sc (q) }\;\bar\chi^{\mu}{\Sc (p_1^2) }
     \end{pmatrix} \; . 
 \label{Kdef}
\end{eqnarray}
Herein, the flavor and color factors have been taken into account explicitly,
and $\chi^5, \, \chi^{\mu}$ stand for the Dirac structures of the
diquark-quark vertices, see eqs.~(\ref{chi5def},\ref{chimudef}). 
The freedom to partition the total momentum between quark and diquark
introduces the parameter $\eta \in [0,1]$ with $p_q=\eta P+p$ and
$p_d=(1-\eta)P - p$. The momentum of the exchanged quark is then given by
$q=-p-k+(1-2\eta)P$. The relative momenta of the quarks in the diquark
vertices  $\chi$ and  $\bar\chi$ are $p_2=p+k/2-(1-3\eta)P/2$ and
$p_1=p/2+k-(1-3\eta)P/2$, respectively.
Invariance under (4-dimensional) translations implies that for
every solution  $\Psi(p,P;\eta_1)$ of the BS equation there exists
a family of solutions of the form $\Psi(p+(\eta_2-\eta_1)P,P;\eta_2)$.

Using the positive energy projector with
nucleon bound-state mass $M_n$,
\begin{equation}
 \Lambda^+= \frac{1}{2}\left( 1+ \frac{\Slash{P}}{iM_n}\right),
\end{equation}
the wave function can be decomposed into their most general Dirac
structures,
\begin{eqnarray}
 \Psi^5(p,P)&=& (S_1 +\frac{i}{M_n}\Slash{p} S_2) \Lambda^+, \qquad
   \label{psi5deco}\\
 \Psi^\mu(p,P)&=& \frac{P^\mu}{iM_n} (A_1 +\frac{i}{M_n}\Slash{p} A_2) \gamma_5
      \Lambda^+  + 
               \gamma^\mu (A_3 +\frac{i}{M_n}\Slash{p} A_4)  
                   \gamma_5\Lambda^+  \label{psimudeco}\\
              &+ & \frac{p^\mu}{iM_n}( A_5 + \frac{i}{M_n}\Slash{p} A_6)
                  \gamma_5\Lambda^+ \; \nonumber .
\end{eqnarray}
In the rest frame of the nucleon, $P = ( \vec 0 , iM_n)$,
the unknown scalar functions $S_i$ and $A_i$ are functions of $p^2=p^\mu
p^\mu$ and of the angle variable $z= \hat P \cdot \hat p$,
the cosine of the (4-dimensional) azimuthal angle of $p^\mu$.
Certain linear combinations of these eight covariant components then
lead to a full partial wave decomposition, see ref.~\cite{Oettel:1998bk} 
for more details and for examples of decomposed amplitudes assuming pointlike
diquarks. Note that such a decomposition in Dirac and Lorentz space
holds for the vertex function $\Phi(p,P)$ as well.

The Faddeev solutions are normalized by the canonical condition
\begin{eqnarray}
  M_n \Lambda^+ \;& \stackrel{!}{=}&
  -\int \frac{d^4\,p}{(2\pi)^4}
  \int \frac{d^4\,p'}{(2\pi)^4} \label{normnuc}\\
   & &\bar \Psi(p',P_n) \left[ P^\mu \frac{\partial}{\partial P^\mu}
    G^{-1} (p',p,P) \right]_{P=P_n} \hskip -.5cm  \Psi(p,P_n) \; .
  \nonumber
 \end{eqnarray}

\subsection{ Delta }

The effective multi-spinor for the delta baryon representing
the BS wave function can be characterized as $\Psi_\Delta^{\mu\nu}(p,P)
u^\nu(P)$ where $u^\nu(P)$ is a Rarita-Schwinger spinor.
The momenta are defined analogously to the nucleon case.
As the delta state is flavor symmetric, only the axialvector
diquark contributes  and,  accordingly, the corresponding BS equation reads,
\begin{eqnarray}
  \fourint{k} G^{-1}_\Delta (p,k,P)
  \Psi_\Delta^{\mu'\nu}(k,P) =0 \; ,
  \label{bse_del}
\end{eqnarray}
where the inverse quark-diquark propagator $G^{-1}_\Delta$ in the
$\Delta$-channel is given by
\begin{eqnarray}
  G^{-1}_\Delta(p,k,P) &=&  (2\pi)^4 \delta^4(p-k)\; S^{-1} (p_q) \;
        (D^{\mu\mu'})^{-1} (p_d) + \nonumber \\
    & &  \chi^{\mu'}(p_2^2)\; S^T(q)\;\bar\chi^\mu(p_1^2).
\end{eqnarray}
The general decomposition of the corresponding vertex function
$\Phi^{\mu\nu}_\Delta$, obtained as in eq.\ (\ref{amp})
by truncating the quark and diquark legs of the BS wave function
$\Psi_\Delta^{\mu\nu}$, reads
\begin{eqnarray}
 \Phi^{\mu\nu}_\Delta (p,P) &=& (D_1 + \frac{i}{M_\Delta} \Slash{p} D_2) \
                        \Lambda^{\mu\nu} + 
     \frac{P^\mu}{iM_\Delta} (E_1 + \frac{i}{M_\Delta} \Slash{p} E_2)
        \frac{p^{\lambda}_{\rm T}}{iM_\Delta} \Lambda^{\lambda\nu} +
        \label{Deldec}\\
   & &  \gamma^\mu (E_3 + \frac{i}{M_\Delta} \Slash{p} E_4 )
        \frac{p^{\lambda}_{\rm T}}{iM_\Delta} \Lambda^{\lambda\nu} + 
     \frac{p^\mu}{iM_\Delta} ( E_5 + \frac{i}{M_\Delta} \Slash{p} E_6)
        \frac{p^{\lambda}_{\rm T}}{iM_\Delta} \Lambda^{\lambda\nu} \; . \nonumber
\end{eqnarray}
Here, $\Lambda^{\mu\nu}$ is the Rarita-Schwinger projector,
\begin{eqnarray}
  \Lambda^{\mu\nu} \!= \Lambda^+
             \left(
             \delta^{\mu\nu}-\frac{1}{3}\gamma^\mu\gamma^\nu
             +\frac{2}{3} \frac{P^\mu P^\nu}{M_\Delta^2} -
             \frac{i}{3} \frac{P^\mu\gamma^\nu-P^\nu\gamma^\mu}{M_\Delta}
              \right)
             \end{eqnarray}
 which obeys the
constraints
\begin{equation}
  P^\mu \Lambda^{\mu\nu} = \gamma^\mu \Lambda^{\mu\nu} =0.
\end{equation}
Therefore, the only  non-zero components arise from the
contraction with the transverse relative momentum
$p^{\mu}_{\rm T}=p^\mu - \hat P^\mu (p\cdot \hat P)$.
The invariant functions $D_i$ and $E_i$ in eq.~(\ref{Deldec}) again depend
on $p^2$ and $\hat p\cdot \hat P$.
The partial wave decomposition in the rest frame is given in
ref.~\cite{Oettel:1998bk}. 

\subsection{Numerical solutions}

The Faddeev equations for $N$ and $\Delta$ are solved in the baryon rest frame
by expanding the unknown scalar functions
in terms of Chebyshev polynomials of the variable $\hat p\cdot \hat P$ 
\cite{Oettel:2002kd}.
Thus the equations are reduced to a system of homogeneous 
one--dimensional integral
equations.  Iterating the integral equations yields a certain eigenvalue
which by readjusting the parameters of the model is tuned to one.
As remarked earlier, we are left with one free parameter which is taken
to be the width $w_{1^+}$ of the axialvector diquark.

\begin{table}[t]
 \begin{center}
 \begin{tabular}{lllllll} \hline \hline
   Set& & I & II & III & IV & V \\ \hline
   $w_{1^+}$ & [GeV$^2$] & 0.4 & 0.6 & 0.8 & 1.0 & 1.2 \\ \hline
   $w_{0^+}$ & [GeV$^2$] & 0.21 & 0.27 & 0.32 & 0.36 & 0.39 \\
   $m_{1^+}$ & [GeV]     & 0.92 & 0.91 & 0.89 & 0.88 & 0.87 \\
   $m_{0^+}$ & [GeV]     & 0.75 & 0.77 & 0.80 & 0.84 & 0.86 \\ \hline \hline
 \end{tabular}
 \end{center}
 \caption{Five parameter sets which describe the physical masses
  of $N$, $M_n=$940MeV, and $\Delta$, $m_\Delta =$1230MeV.} 
 \label{parsets}
\end{table}

In Tab.~\ref{parsets} we show five parameter sets which lead to a bound nucleon
and delta with the correct physical masses, $M_n=0.94$ GeV 
and $M_\Delta=1.23$ GeV. Instead of the $\Lambda$ parameters the
pole locations  in the diquark propagators are tabulated 
which have a more intuitive interpretation. 

We need always a broader axialvector diquark (in momentum space)
to fit both nucleon and delta. The resulting diquark masses, notably
the mass difference $m_{1^+}-m_{0^+}$, agrees approximately with previous
rainbow--ladder results \cite{Praschifka:1989fd,Burden:1997nh}
only for the first two sets with smaller diquark widths.
The lattice results of ref.~\cite{Wetzorke:2000ez} give 
$m_{0^+}=0.83$ GeV and $m_{1^+}=0.9$ GeV, within the spread of our
parameter sets.

As mentioned in the introductory chapter, pions are expected to reduce
the nucleon mass by at least 200 MeV \cite{Hecht:2002ej}. 
The corresponding mass shift
for the delta will be lower. Therefore we investigate the Faddeev
equations for nucleon and delta core states of mass 
1.2 and 1.4 GeV respectively and present the results in
Appendix~\ref{heavy-sec}.

\section{Nucleon form factors}
\label{ff}

\subsection{Electromagnetic gauge invariance}

To calculate form factors, we apply the gauging formalism
of ref.~\cite{Kvinikhidze:1999xn} which basically consists in coupling
the photon to all elements in the kernel $G^{-1}$ of the nucleon
Faddeev equation (\ref{bse_nuc}). Therefore we need the photon vertices
with quark, diquarks and the quark exchange kernel. Each vertex has to
satisfy its WT identity. 

For the quark--photon vertex 
$\Gamma^\mu_q=\Gamma^\mu_{q,{\rm BC}}+\Gamma^\mu_{q,{\rm T}}$ 
the construction of the longitudinal
part, $\Gamma^\mu_{q,{\rm BC}}$, which is fixed by the WT identity 
has been long known \cite{Ball:1980ay}.
It is given by
\bea
 \label{qvertex}
 \Gamma^\mu_{q,{\rm BC}}(k,p) = -i\gamma^\mu \frac{A_k+A_p}{2}-i(p+k)^\mu
 \frac{\Slash{k}+\Slash{p}}{2} \Delta A-
 (p+k)^\mu \Delta B
\eea
where $\Delta X=(X_k-X_p)/(k^2-p^2)$ and $X_k=X(k^2)$, $(X=\{A,B\})$. The
remaining transverse part, $\Gamma^\mu_{q,{\rm T}}$ is yet undetermined. 
To ensure multiplicative 
renormalizibility at the one--loop level, an {\em ansatz} for this part
was proprosed in ref.~\cite{Curtis:1990zs} but it modifies our results for 
form factors only on the level of one per cent. The transverse part might also
receive dynamical contributions from the $\rho-\omega$ meson poles in the
$q-\bar q$ vector channel \cite{Maris:2000bh}. 
In Appendix~\ref{vert_res}, we derive a parametrization of such 
ontributions which is well--constrained
by the {\em pion} form factor.
It is given by
\bea
 \label{res_v}
 \Gamma^\mu_{q,{\rm T}}(k,p)&=& \phi^\mu\;
  \frac{m_\rho}{f_\rho} \frac{Q^2}{Q^2+m_\rho^2}\;e^{-\alpha\left(
  1+\frac{Q^2}{m_\rho^2}\right)} \; , \\
 \phi^\mu& = &\left(i\gamma^\mu_{\rm T} - 1.69 \frac{q^\mu_{\rm T}}{\omega_\rho}
   \right)\; \frac{{\cal F}^2(q^2/\omega^2_\rho)}{0.139} \; , \\
  Q &=& k-p \; , \qquad q=(k+p)/2\; \qquad 
  v^\mu_{\rm T}=v^\mu -Q^\mu v\cdot Q /Q^2 \nonumber \; .
\eea
The constants appearing herein are: $\rho$ mass and decay constant
$m_\rho=0.77$ GeV and $f_\rho=0.215$ GeV (calculated), $\alpha=0.652$
and $\omega^2_\rho=0.35$ GeV$^2$. 
The structure $\phi^\mu$ 
represents a properly normalized
vector meson BS amplitude (see Appendix~\ref{vert_res}).
In eq.~(\ref{res_v}), the combination of the exponential and the 
propagator--like denominator parametrizes the effects of the
propagation of an off-shell, composite $\rho-\omega$. For details, see
ref.~\cite{Maris:2000bh}.

\begin{figure}
 \begin{center}
   \epsfig{file=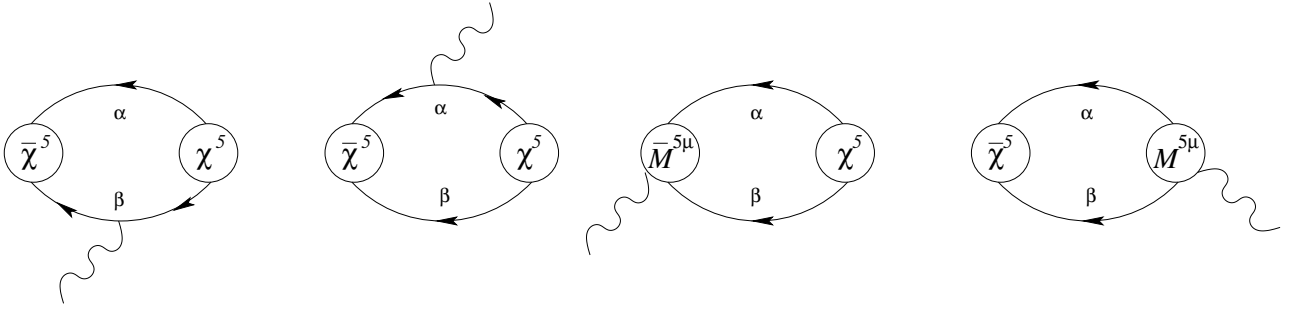,width=\columnwidth}
 \end{center}
 \caption{The photon--diquark vertex, here for the scalar diquark.
For the axialvector diquark, $\chi^5$ and $M^{5\mu}$ have to be replaced
by their corresponding counterparts. Note that in actual numerical
calculations the first two diagrams are equivalent if the quarks are identical.}
 \label{dqv_f}
\end{figure}

The diquark--photon vertices $\Gamma^{\mu}_{0^+}$ and
$\Gamma^{\mu}_{1^+}$ receive contributions from four different
diagrams, depicted in Fig.~\ref{dqv_f}. Besides the photon coupling to the
quarks within the loop we need seagull graphs which describe the photon
coupling to the vertices $\chi$ and $\chi^\mu$ to recover the WT identity
$(k-p)^\mu \Gamma^{\mu}_{0^+[1^+]}=(D^{[\mu\nu]})^{-1}(k) - 
(D^{[\mu\nu]})^{-1}(p)$. The functional form of these seagull vertices 
has been derived in ref.~\cite{Oettel:2000gc} and they read
\begin{eqnarray}
 \label{seagull}
  (M^a)^\mu(p',Q;q_\alpha,q_\beta)&=& q_\alpha \;\frac{(4p'-Q)^\mu}{4p'\cdt Q-Q^2}
    \;\left[ \chi^a(p'-Q/2) -\chi^a(p') \right] + \nonumber \\
  & &  q_\beta\; \frac{(4p'+Q)^\mu}{4p'\cdt Q+Q^2}
     \;\left[ \chi^a(p'+Q/2) -\chi^a(p') \right] \; . \label{seag1}
\end{eqnarray} 
The photon momentum is denoted by $Q=k-p$. The relative momentum $p'$ between
the two quarks with charges $q_\alpha$ and $q_\beta$ has been defined in
eq.~(\ref{dq_rm}). The conjugated vertex is obtained by replacing
$\chi \to \bar \chi$, $Q \to -Q$ and interchanging $q_\alpha \leftrightarrow
q_\beta$.

Photon--mediated transitions between scalar and axialvector diquarks 
are also possible. The corresponding (anomalous) vertices $\Gamma_{0^+-1^+}$
resp. $\Gamma_{1^+-0^+}$ are described by diagrams like the first two
in Fig.~\ref{dqv_f}. Seagulls give no contributions to these vertices.

\begin{figure}
 \begin{center}
   \epsfig{file=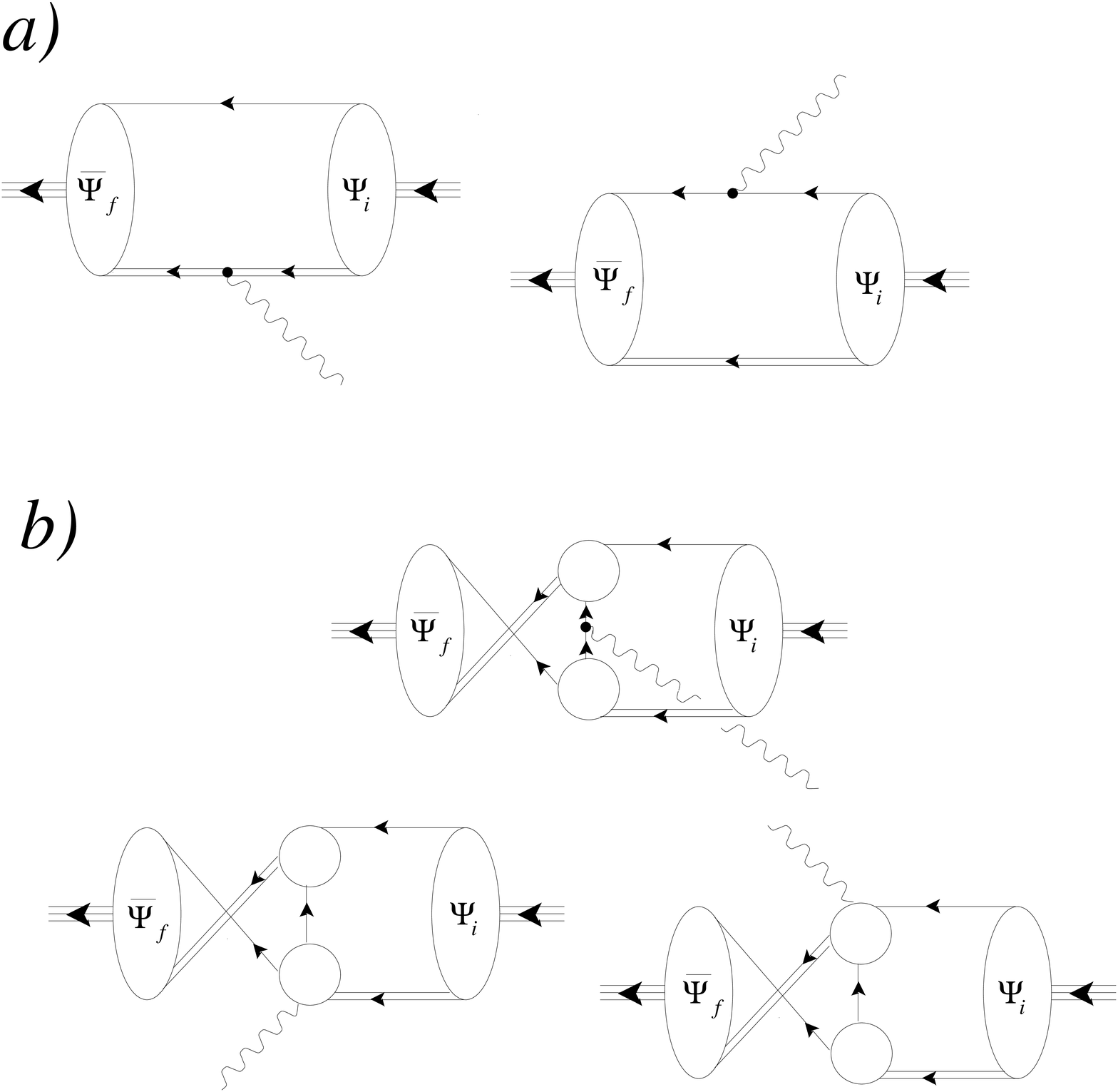,width=10cm}
 \end{center}
 \caption{Nucleon current matrix elements: 
 (a) Impulse approximation diagrams. The photon--diquark vertex consists
     of the elements given in Fig.~\protect{\ref{dqv_f}}. (b) Exchange kernel
     diagrams.}  
 \label{ff_f}
\end{figure}

The quark exchange kernel, as given in eq.~(\ref{Kdef}), consists
of expressions \\ 
$\sim \chi^a\; S^T \; \bar \chi^b$ ($a,b=\{5,\nu\}$). 
Complete gauging leads to a diagram where the photon couples to the
exchange quark and two diagrams where the photon couples 
to the vertices $\chi^a$ and $\bar \chi^b$. The latter couplings
are described by the vertex as given in eq.~(\ref{seagull}).
The proof that the gauged quark exchange kernel obeys its WT identity
can be found in ref.~\cite{Oettel:2000ig}.

In summary, to obtain the complete nucleon current matrix element, 
we have to calculate the diagrams shown in Fig.~\ref{ff_f}.  
We remark in passing that the normalization condition for the
nucleon Faddeev amplitudes, eq.~(\ref{normnuc}), is only compatible with
the correct nucleon charges, i.e. $G_E(0)=1$ (proton) and $G_E(0)=0$
(neutron), if {\em all} diagrams of Fig.~\ref{ff_f} are taken into
account. 

\subsection{Numerical calculations}

We extract the Sachs electromagnetic form factors from the current matrix
elements by the following traces:
\begin{eqnarray}
 \label{getrace}
G_E(Q^2) &=& \frac{M_n}{2P^2}\; {\rm Tr}\;
    \langle P_f| \;  J^\mu\; |P_i \rangle  P^\mu \; ,\\
 \label{gmtrace}
G_M(Q^2) &=& \frac{iM_n^2}{Q^2}\; {\rm Tr}\;\langle P_f| \;  J^\mu\; |P_i \rangle
   (\gamma^\mu)_{\rm T} \; ,\qquad
   \left( (\gamma^\mu)_{\rm T}=\gamma^\mu -\hat P^\mu \hat{\Slash{P}}\right)\; .
  \quad
\end{eqnarray}
Here, $P=(P_i+P_f)/2$, and the initial and final states $|P_i\rangle$ and
$\langle P_f|$ are given by the numerical solutions for the matrix valued
wave functions $\Psi(p,P_i)$ and $\bar \Psi(k,P_f)$, {\em cf.} 
eqs.~(\ref{psidef},\ref{psi5deco},\ref{psimudeco}).

Due to the complicated singularity structure of the single 
diagrams and due to limited computer resources we obtained fairly accurate
numbers for the form factors only up to momentum transfers of $Q^2=2$ GeV$^2$.
These technical obstructions do not interfere with the conclusions we will
draw, though. A detailed discussion of the technicalities is deferred
to Appendix~\ref{tech}.

\begin{figure}
\begin{center}
 \epsfig{file=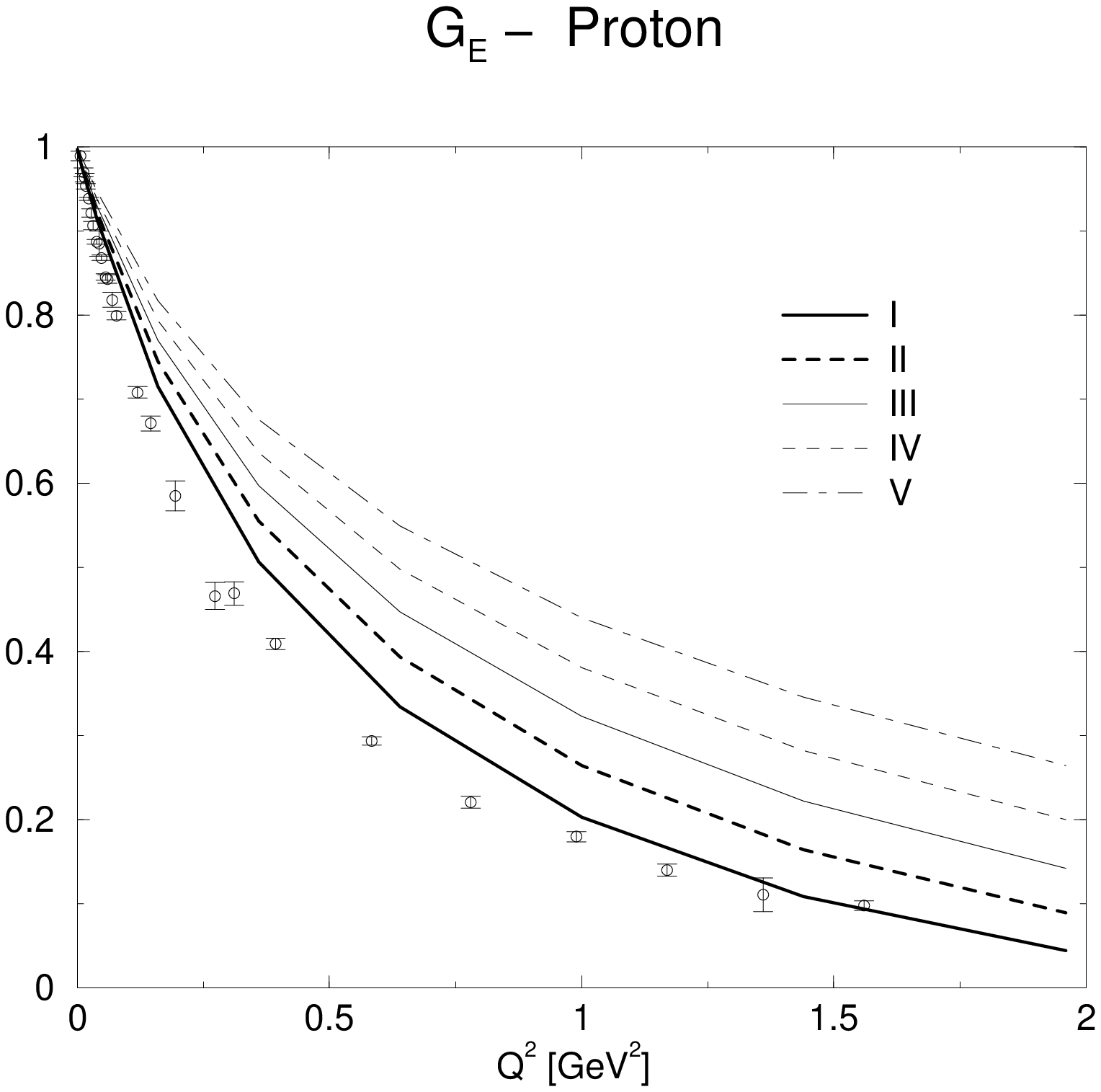,width=6cm} \hskip 2cm
 \epsfig{file=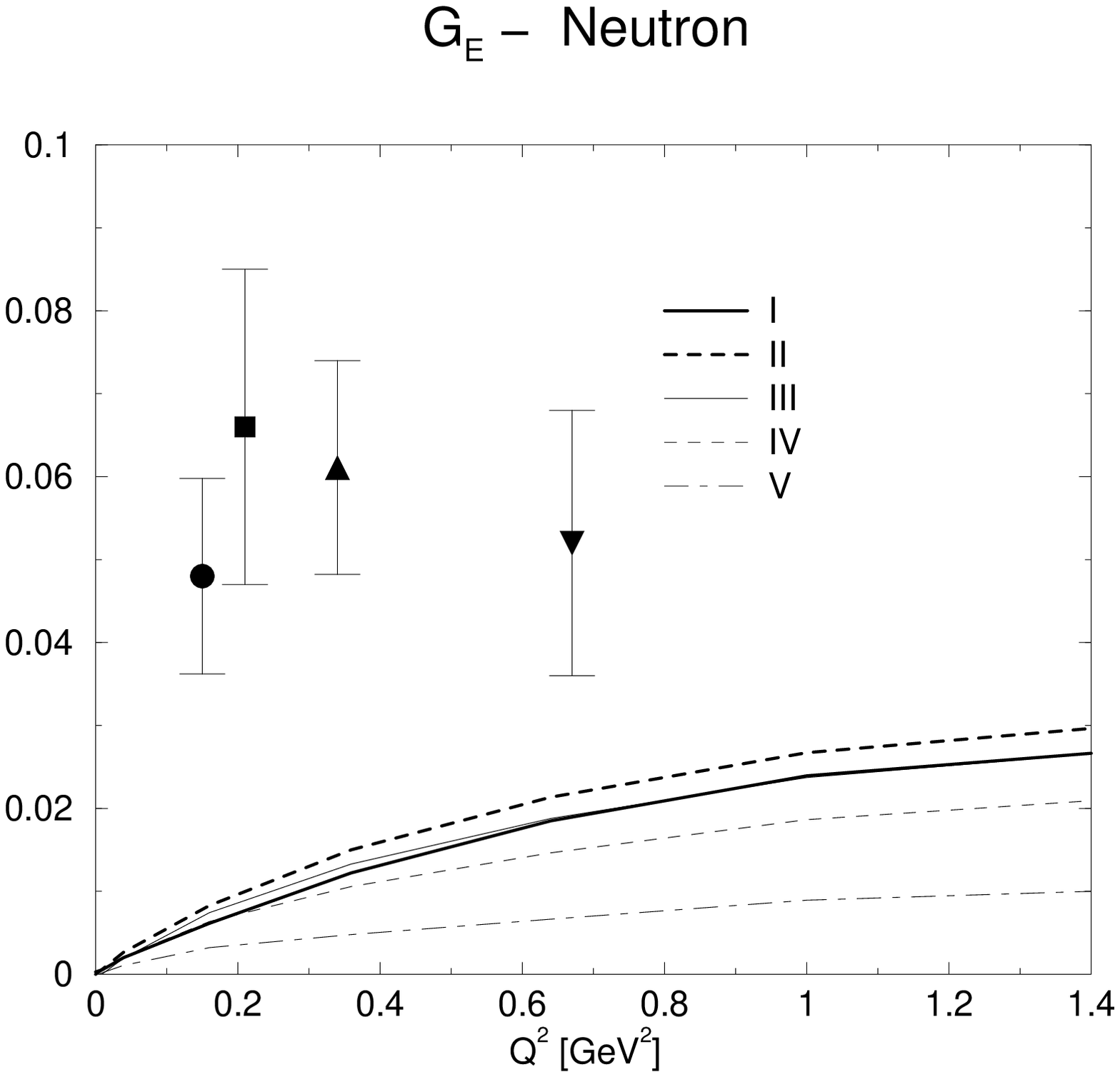,width=6cm}
\end{center}
 \caption{Depicted are the electric form factors of proton and neutron. 
  Experimental data for the proton are
 taken from the analysis in ref.~\cite{Hoehler}. 
 For the neutron, data are from ref.~\cite{Herberg:1999ud} (diamond),
 from ref.~\cite{Passchier:1999cj} (square), from ref.~\cite{Ostrick:1999xa}
 (triangle up) and finally from ref.~\cite{Rohe:1999sh} (triangle down).}
 \label{ge_f}
\end{figure}

We have calculated the form factors for the five sets tabulated in 
Tab.~\ref{parsets}. The results for the electric form factors
of proton and neutron are shown in Fig.~\ref{ge_f}. 
The proton $G_E$ (left panel) becomes steeper with
decreasing diquark widths $w_{1^+}$ and $w_{0^+}$. This is in agreement with
intuition since the scalar diquark correlations give the most important
contributions to this form factor, and with decreasing width also 
$m_{0^+}$ decreases,
 so that these correlations become wider in the 
``center--of--mass'' position variable as in the relative position variable. 
However, we also observe for Set I an interesting deviation of the form factor from 
the dipole shape which will be discussed below. 

\begin{table}[b]
 \begin{center}
 \begin{tabular}{llrrrrr} \hline \hline
   Set& & I & II & III & IV & V \\ \hline
  $\mu_p$& [n.m.] & 3.05    & 2.94    & 2.86    & 2.79    & 2.73 \\
  $\mu_n$& [n.m.] & $-$1.78 & $-$1.65 & $-$1.55 & $-$1.47 & $-$1.40 \\
 \hline \hline
 \end{tabular}
 \end{center}
 \caption{The magnetic moments of proton ($\mu_p$) and neutron
   ($\mu_n$) for the five parameter sets given in table 1.}
 \label{mu_t}
\end{table}

Let us turn to the results for $G_E$ of the neutron (right panel 
in Fig.~\ref{ge_f}). All data
sets predict a positive form factor  which is  slowly falling for larger $Q^2>1$
GeV$^2$.
No data set can  reproduce the experimental neutron charge radius 
or come close to it. 
This is in remarkable contrast to the results in 
refs.~\cite{Oettel:2000jj,Bloch:1999ke} where rather simple parametrizations
of the $q-q$ $t$ matrix were employed. As for ref.~\cite{Oettel:2000jj},
the good description of $G_E$ was mainly a result of the cancellation
between quark and diquark impulse approximation diagram. The former
contribute negatively, the latter positively and by virtue of
the simple approximation of the $q-q$ $t$ matrix by free spin--0/spin--1
particles and the corresponding free photon vertices, the diquark 
contributions fall slower and thus render $G_E$ positive.
In this study, we have resolved the diquarks (see Fig.~(\ref{dqv_f}))
and the effect of diquark contributions falling more slowly is almost
absent, thus the charge radius becomes very small. Only with a proper 
resolution of the diquarks an asymptotically correct description for
the form factors is possible at all, thus we conclude that the
neutron charge radius must be accounted for by other mechanisms
such as a neutron dressing by pions. 
Nevertheless the positivity of $G_E$ for higher
momentum transfers is a result of the fully relativistic treatment.

\begin{figure}
\begin{center}
 \epsfig{file=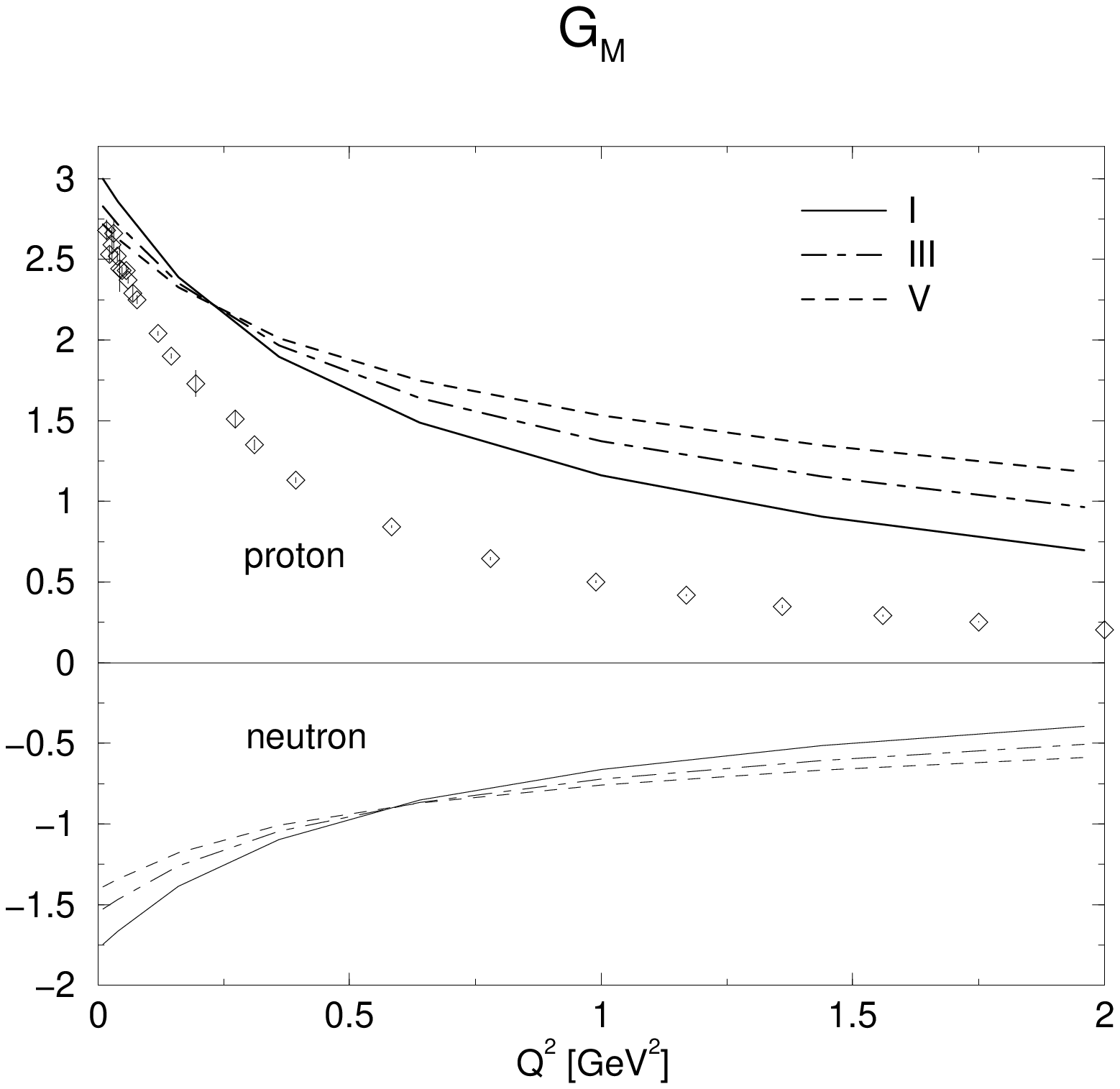,width=6cm} \hskip 2cm
 \epsfig{file=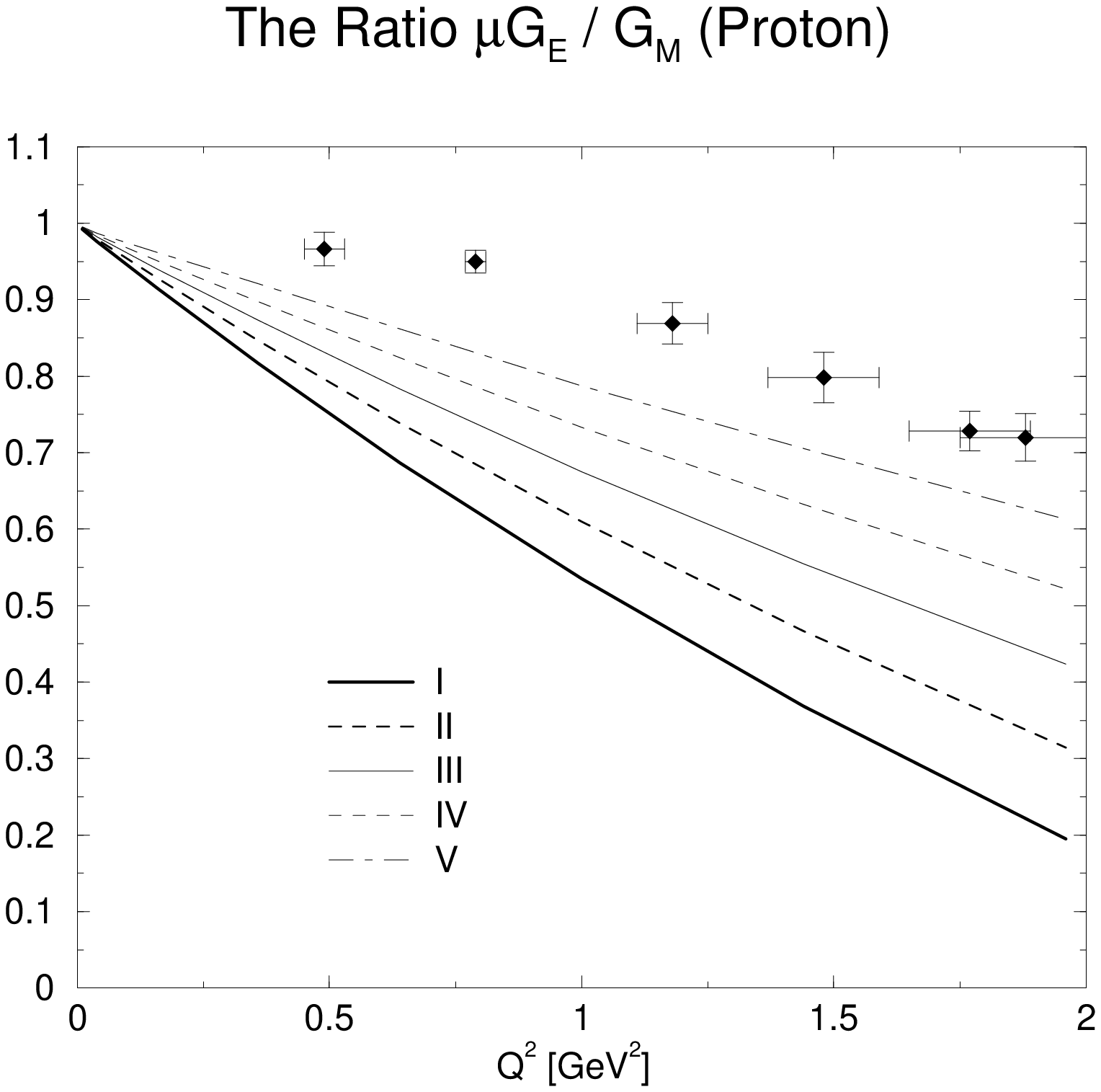,width=6cm}
\end{center}
 \caption{ Magnetic form factors and $G_E/G_M$. Experimental data for
 proton's $G_M$ are from ref.~\cite{Hoehler} and data for
 $G_E/G_M$ have been reported in ref.~\cite{Jones:2000rz}.} 
 \label{gm_f}
\end{figure}

Fig.~\ref{gm_f}, left panel, shows the nucleon magnetic form factors 
and Tab.~\ref{mu_t} the corresponding magnetic moments. Also the magnetic radii
become larger with decreasing diquark widths, as well as $\mu_p$ and 
$|\mu_n/\mu_p|$. 

Our results for the ratio $\mu_p G_E/G_M$, currently under intensive 
experimental scrutiny, are shown in the right panel of Fig.~\ref{gm_f}.
Although the results consistently put that ratio below 1, the available 
experimental data are underestimated considerably. Even worse, as
going towards more realistic electromagnetic radii (with decreasing 
set number) the ratio becomes smaller and smaller. Before giving
a reason for the underestimation of $\mu_p G_E/G_M$, we will examine the
influence of the vector mesons in the quark--photon vertex.

{\em Full vertex vs. Ball--Chiu vertex}

Since the resonance contribution is $\sim Q^2$ near $Q^2=0$, it does
not give any contributions to the magnetic moments, thus these are accounted
for by the Ball--Chiu vertex alone. As for the pion form factor, to which
the resonant vertex was fitted, it does give sizeable contribution to the 
charge radii. It amounts to 21--23 \% in the case of 
$(r_p)^2_{\rm el}$ for all data sets and is therefore of the same relative 
size as the contribution of the resonance to the pion charge radius, 
see Appendix~\ref{vert_res}. The neutron charge radius is usually a bit smaller,
as the electric form factor is quenched a bit more when the full vertex is 
employed, nevertheless the differences are small and can not account for
the discrepancy with the data. 

\begin{figure}
\begin{center}
 \epsfig{file=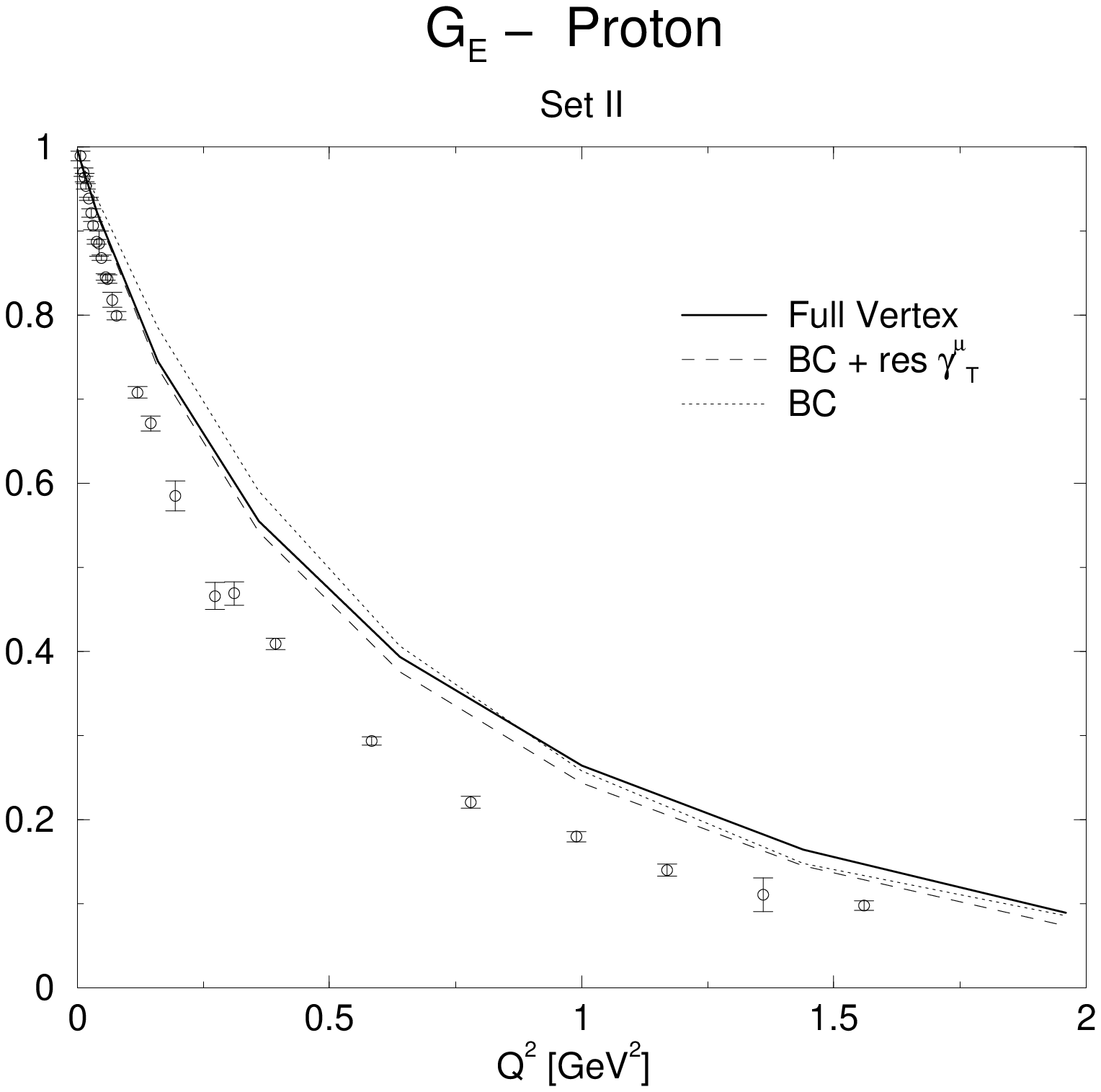,width=6cm} \hskip 2cm
 \epsfig{file=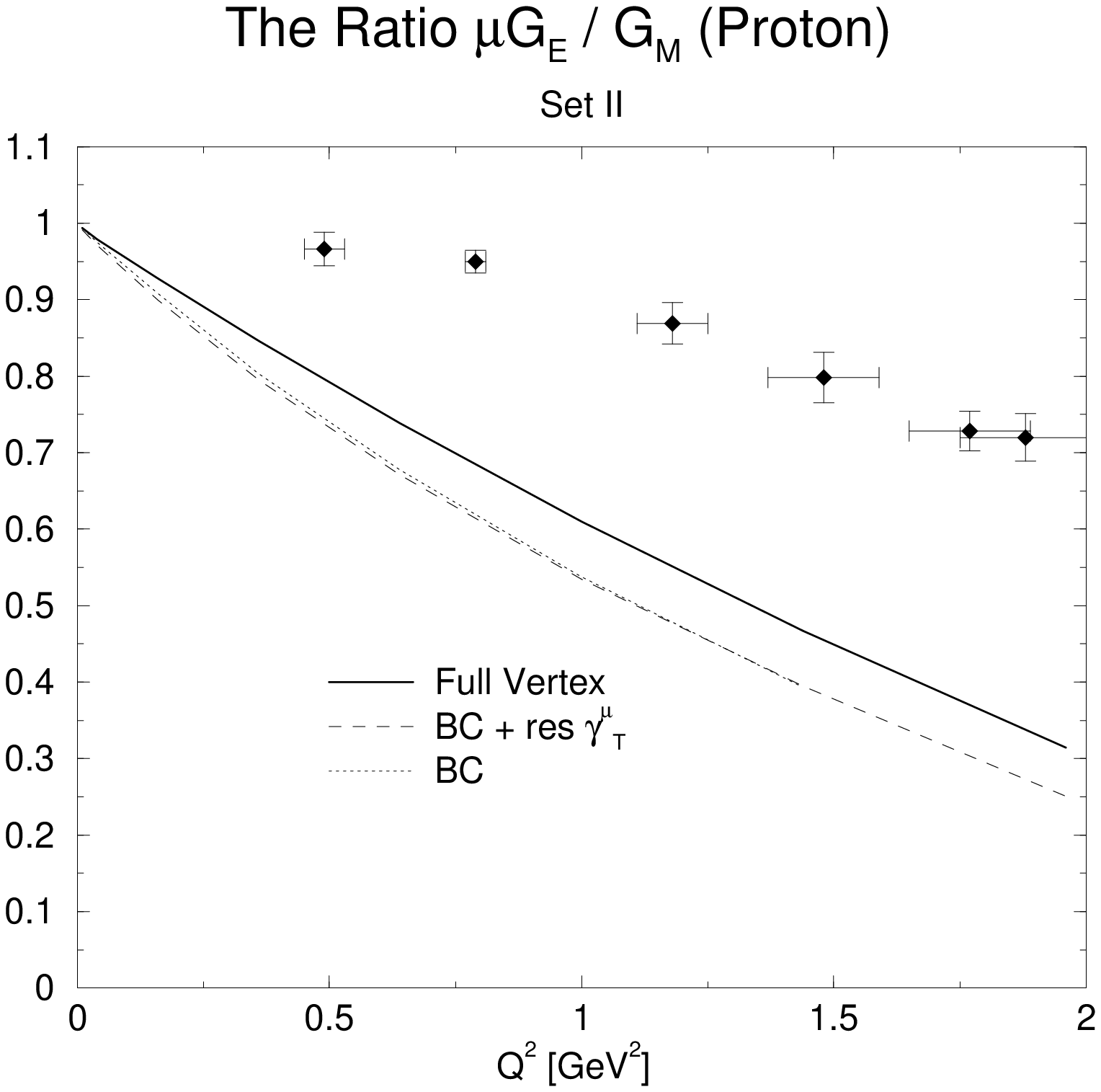,width=6cm}
\end{center}
 \caption{ Comparison between results for the full vertex, the
 Ball--Chiu (BC) vertex plus a dominant resonance term, and the pure
 Ball--Chiu vertex. In the first two cases, the resonance terms have been 
 fitted to the pion form factor.} 
 \label{gcomp_f}
\end{figure}

In Fig.~\ref{gcomp_f} we show the influence of the resonance term on 
proton's $G_E$ and the ratio $G_E/G_M$ for set II. We investigated two cases
of resonance contributions, the full vertex which includes
the transverse terms as in eq.~\ref{res_v} and a transverse vertex
which includes only the leading $\gamma^\mu_{\rm T}$ term. In both cases,
the vector meson amplitude is normalized and the damping constant $\alpha$
has been fitted to the pion form factor. For small $Q^2$, both resonance
parametrizations lead to a $G_E$ falling more quickly, with almost no 
quantitative difference. For intermediate $Q^2$, the subleading
vector meson amplitude $\sim q_{\rm T}$ leads to a slightly enhanced 
$G_E$. This effect is visible more clearly in the ratio $G_E/G_M$. 
The subleading amplitude also quenches $G_M$ a bit such that the
ratio comes out more than 10 \% larger than for the BC vertex in
the intermediate $Q^2$ domain.

We see that the vector meson resonance has a sizeable influence
on the proton charge radius (of the order 1/4), and subdominant
amplitudes of the vector meson can influence $G_E/G_M$ 
by 10--15 \%. Nevertheless for this observable a discrepancy remains,
and a reason
 can be found by analyzing the Ball--Chiu
quark--photon vertex. This vertex always appears with quark legs attached,
i.e. in the combination $\tilde\Gamma^\mu_{q} =  
S(k)\Gamma^\mu_{q,{\rm BC}} S(p)$.
We rewrite $\tilde\Gamma^\mu_q$ in the following way:
\begin{equation}
 \tilde\Gamma^\mu_q = \tilde\Gamma^\mu_{q,{\rm BC}} + \tilde\Gamma^\mu_{\rm T}\; . 
\end{equation}
The vertex with legs fulfills the WT identity
$Q^\mu \tilde\Gamma^\mu_q =S(p)-S(k)$ (note that the propagator and not its
inverse appears on the r.h.s.). The term $\tilde\Gamma^\mu_{q,{\rm BC}}$
is constructed via the Ball--Chiu technique 
to satisfy this identity:
\begin{equation}
 \tilde\Gamma^\mu_{q,{\rm BC}}(k,p) = -i\gamma^\mu \frac{\sigma_{Vk}+\sigma_{Vp}}{2}
  -i(p+k)^\mu
 \frac{\Slash{k}+\Slash{p}}{2} \Delta \sigma_V+
 (p+k)^\mu \Delta \sigma_S \; .
\end{equation}
The remainder, $\tilde\Gamma^\mu_{\rm T}$, is transversal, and after some
Dirac algebra one finds
\bea
  \tilde\Gamma^\mu_{\rm T} &=& \left[ Q^\mu - \gamma^\mu \Slash{Q} +
    i\Delta M (\gamma^\mu (k^2-p^2)-\Slash{Q}(k+p)^\mu)\right]
    \frac{i\Slash{p}-M_p}{p^2+M_p^2} \frac{\sigma_{Vk}}{2}  -  \nonumber\\
 &&\frac{\sigma_{Vp}}{2} \frac{i\Slash{k}-M_k}{k^2+M_k^2}
    \left[ Q^\mu - \Slash{Q} \gamma^\mu +
    i\Delta M (\gamma^\mu (k^2-p^2)-\Slash{Q}(k+p)^\mu)\right]\; .
  \label{dMterms}
\eea
Due to the running mass function, the terms proportional
to $\Delta M= (M(k^2)-M(p^2))/(k^2-p^2)$ are non--zero. Precisely
these terms give a fairly large negative contribution to the proton's
$G_E$, thus causing a deviation from the dipole shape. This effect is absent
for $G_M$. Of course, a dynamic quark mass function of the
kind depicted in Fig.~\ref{qprop_f} is phenomenologically
required
 and
thus, as in the case of the neutron electric form factor, the effect
of other contributions besides the quark valence core should be sizeable
for this observable. Nevertheless, some parts of the discrepancy 
(though not all) could
be attributed to a stronger influence of the subdominant vector meson 
amplitudes. This amounts to a shift of the off--shell vector meson 
contributions between the Ball--Chiu vertex and the transverse
contributions and thus between the quark propagator and the
off-shell vector meson amplitudes. 

\begin{figure}
 \begin{center}
   \epsfig{file=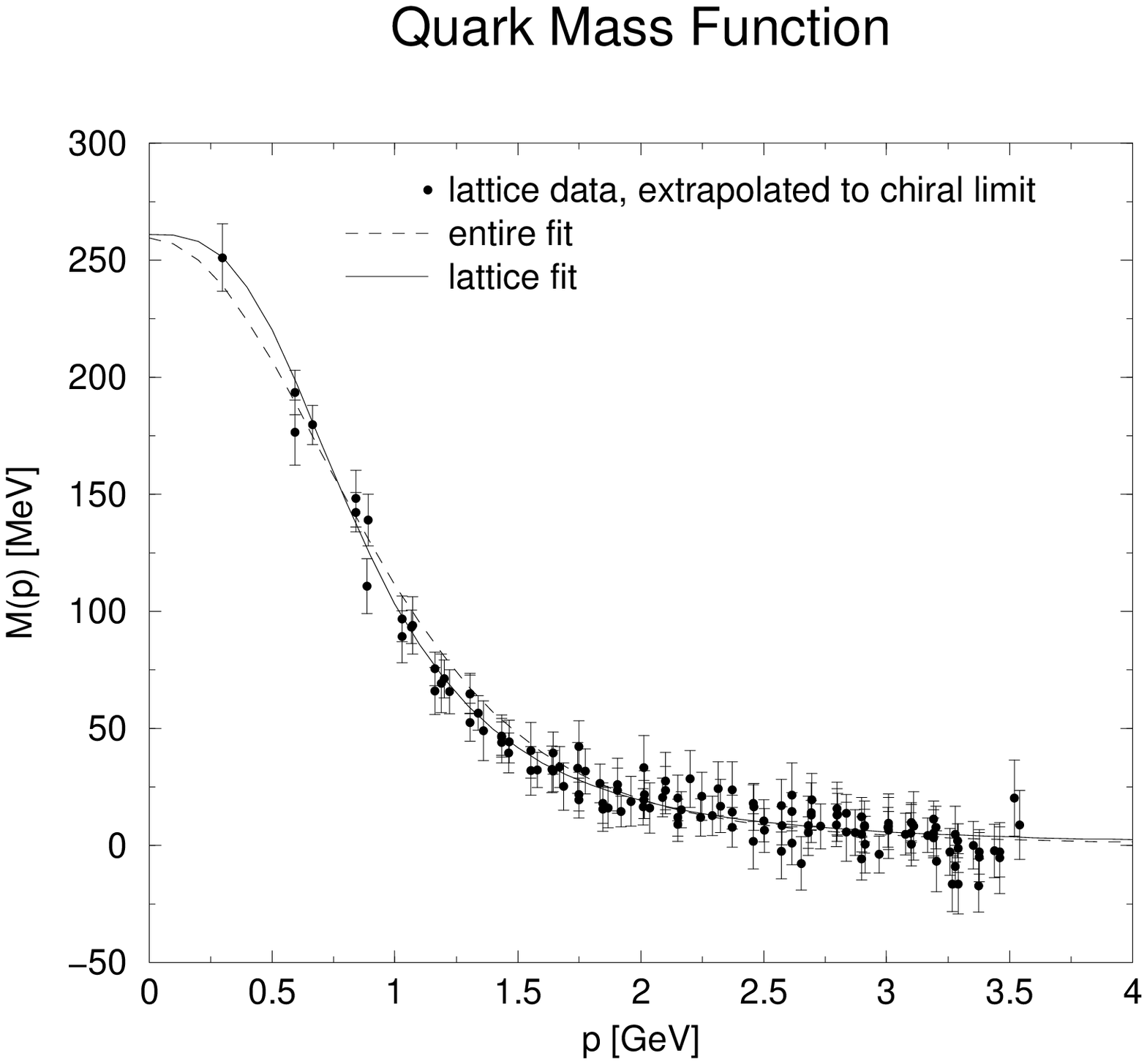,width=6cm} \hskip 2cm
   \epsfig{file=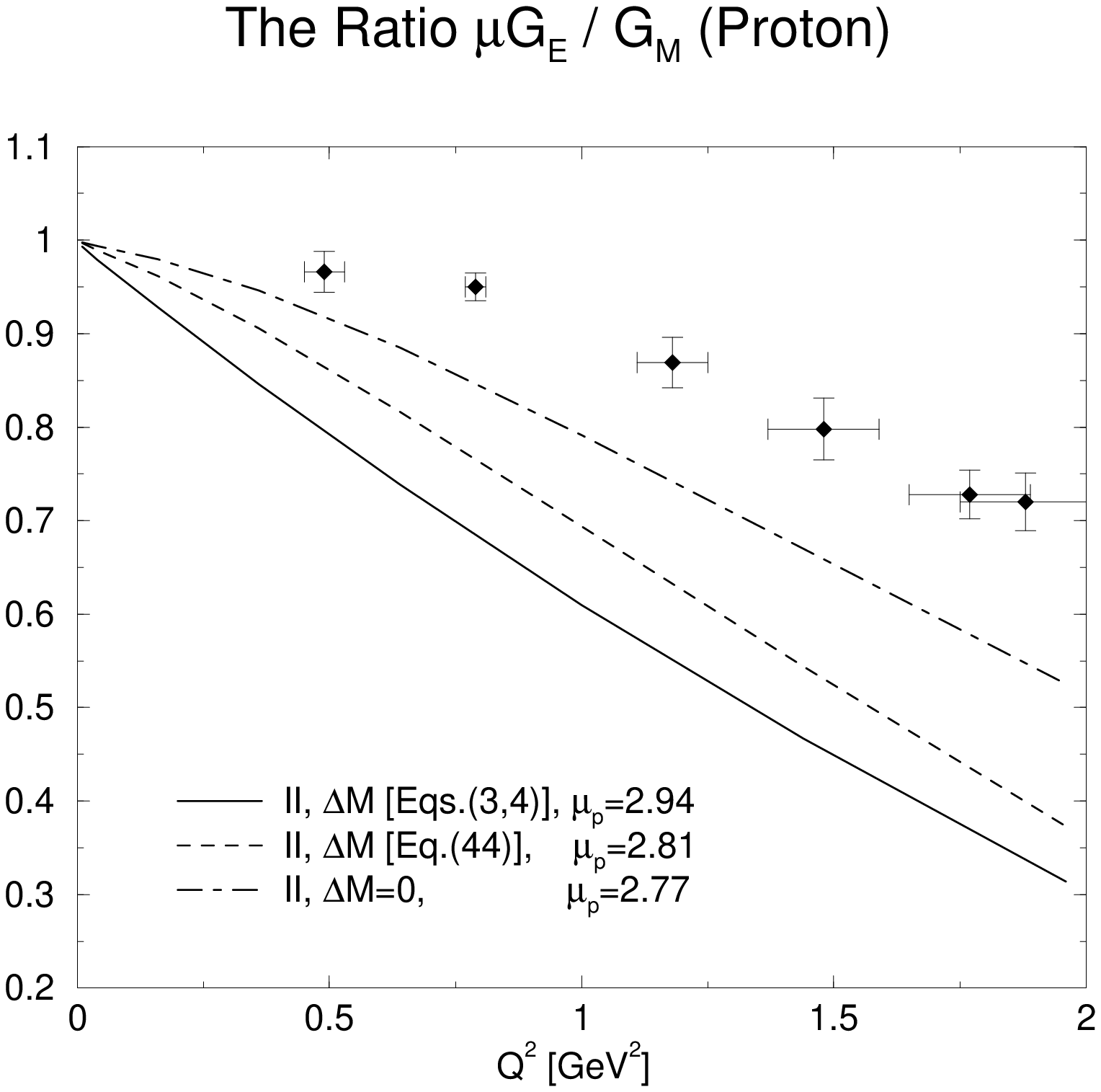,width=6cm}
 \end{center}
 \caption{Left panel: lattice data from ref.~\cite{Bowman:2002bm} 
  for the quark mass function, extrapolated to the chiral limit. 
 Right panel: the ratio $G_E/G_M$ for calculations with modified
 transverse part of the quark--photon vertex.}
 \label{dm_fig}
\end{figure}

To get a first estimate how a change in the quark mass function 
might influence the ratio $G_E/G_M$, we recalculated the form factors
using the wave functions from parameter set II, but replaced all occurences
of $\Delta M$ in the expression for $\tilde\Gamma^\mu_q$, eq.~(\ref{dMterms}),
by (a) zero (corresponding to a momentum--independent quark mass), and
(b) by a fit to the most recent lattice data~\cite{Bowman:2002bm}.
Since we are using wave functions calculated with the quark propagator
parametrization from eqs.~(\ref{ssm},\ref{sigv}), this procedure
is somewhat inconsistent, but may give qualitative indications to the
behavior of $G_E/G_M$. Nevertheless, gauge invariance remains intact
since the $\Delta M$ terms only appear in the transversal part of the vertex. 
We remarked earlier that the quenched--QCD (Landau gauge) 
lattice data seem to suggest
a somewhat broader quark mass function than the parametrization
employed here. This can be seen from Fig.~\ref{dm_fig} (left panel) 
where chiral limit extrapolations of the data and a fit
from ref.~\cite{Bowman:2002bm} are given. Since the functional form
used in the lattice fit, 
$M(p^2)=c\Lambda^{1+2\alpha}/(p^{2\alpha}+ \Lambda^{2\alpha})$ with 
$\alpha=1.52$, has a cut along the negative $p^2$ axis, it is not well suited
for our calculations. We chose to re--fit the data to the entire function
\bea
 \label{fit_anal}
  M(q^2) &=& c_1\left(\frac{1-\exp(-q^2-c_2^2)}{q^2+c_2^2}\right)^2 \; ,
  \quad [q^2=p^2/(1\;{\rm GeV}^2)].
\eea
For the choice of parameters  $c_1=0.4$ GeV and $c_2^2 = 0.45$ 
our fit is also depicted in  Fig.~\ref{dm_fig}. We remark here that
both fits, lattice and entire one, lead to a 20 \% underestimation of 
the pion decay constant $f_\pi$ 
(using the formula from ref.~\cite{Burden:1996ve}), reflecting the 
uncertainties in lattice extrapolations and the shortcomings of the
quenched approximation.

Results for the ratio $G_E/G_M$ are plotted in Fig.~\ref{dm_fig}
(right panel). The ratio is lowest for the consistent calculation.
If eq.~(\ref{fit_anal}) is used in the vertex $\tilde\Gamma^\mu_q$, the
curve is shifted upwards and the proton magnetic moment is somewhat smaller.
Surprisingly the simple constituent quark assumption $\Delta M=0$
delivers the best results compared to the experimental data, though
employing the approximate wave functions prevents us from drawing precise
quantitative conclusions. Nevertheless we see that the ratio $G_E/G_M$
{\em is} sensitive to the precise form of the running quark mass.
Furthermore we note that nearly the
whole effect comes from the quark impulse approximation diagram (the 
second one in Fig.~\ref{ff_f}). 

Clearly more precise QCD lattice data
for the quark propagator and/or DSE/BSE studies of the three--quark and
quark--photon systems are desirable.

Form factor results for nucleons with higher quark--diquark core mass are 
presented in Appendix~\ref{heavy-sec}. There we find that the neutron electric
form factor does not change, and that the even steeper mass function
in the time--like domain deteriorates the ratio $G_E/G_M$ even further.

\section{Summary and conclusions}
\label{concl}

In a step towards the solution of the full covariant Faddeev equations for 
baryons, we have modeled two--quark correlations by 
assuming them separable and by summing quark polarization loop diagrams.
In this case, the Faddeev equations reduce to a Bethe--Salpeter
equation which has been solved exactly.
The technique employed in calculating the two--quark correlations
effectively reduced the number of model parameters to one, the diquark width.

The nucleon form factors have been calculated in a scheme which preserves
the Ward--Takahashi identities for the basic two--point function, 
the quark propagator, for the four--point function, the
quark--quark scattering kernel and finally for the quark--diquark kernel
of the Faddeev equations. Consequently the current is conserved.
Constrained by the pion form factor,
effects from vector mesons have been included in the quark--photon vertex.

Results reveal two effects.
If the proton electric and magnetic radius is to be described 
correctly, the ratio $G_E/G_M$ is severely underestimated. This is a
consequence of the parametrization of the dynamic mass function
of quarks in accordance with results from Dyson--Schwinger and lattice studies.
Furthermore the substructure of the two--quark correlations which is
resolved by the photon renders the neutron form factor positive, but
quite consistently for all parameter sets
the corresponding charge radius cannot be described.
Assuming core nucleon states with higher mass does not alter the above  
findings.

Certainly the precise shape of the form factors is expected to vary
if the technical simplifications can be rendered obsolete, such as the 
separability of the two--quark 
correlations and the treatment of the vector meson contributions
to the quark--photon vertex.
Nevertheless 
it seems possible that the qualitative features will remain valid,
i.e.\ the quenched neutron electric form factor for a correctly resolved
$q-q$ matrix and the underestimation of $G_E/G_M$ due to the quark--photon
vertex with running mass function. As described, the vector meson
contributions cannot compensate this effect. The running quark mass 
and vector mesons are usually not considered in non-- or semirelativistic 
quark models and urge us to a cautious interpretation of corresponding results,
see e.g.\ ref.~\cite{Boffi:2001xv}. 
Thus the investigation presented in this paper 
point towards the necessity to incorporate non--valence
quark physics into the description. Covariant studies of the effect of
e.g. the pion cloud within covariant bound--state perturbation theory
are clearly desirable.

\section*{Acknowledgements}

We thank Craig Roberts for useful discussions in the early stages 
of this project. \\
M.O.\ wants to thank Tony Thomas particularly for a critical
reading of the manuscript and helpful remarks. 
He is grateful for a shared grant by 
the Alexander-von-Humboldt foundation and the CSSM, Adelaide.  \\
R.A.\ is grateful to the members of the CSSM, Adelaide, for their 
hospitality.  \\
This work has been supported by COSY (contract no.\ 41376610).

\begin{appendix}

\section{Results for core nucleon and delta states with higher mass}
\label{heavy-sec}

Pion cloud corrections will lower the mass of the nucleon by more than 200 
MeV. The corresponding mass shift of the delta will be somewhat lower
as the nucleon--delta mass difference is partly also a consequence of
pionic dressing. For exploring effects of this shift on observables,
we fix the core masses to $M_{n,\rm c}=1.2$ GeV and 
$M_{\Delta,\rm c}=1.4$ GeV. Possible parameters which lead to a solution of
the Faddeev equations are given in Tab.~\ref{parsets2}. We note
that less than half of the mass difference between core and physical 
nucleon state can be attributed to the shift in the diquark masses which are
about 100 MeV larger than those in Tab.~\ref{parsets}.
If one assumed pointlike diquarks, the scalar diquark would not be changed
by pionic corrections at all, but the quark substructure allows for
pion dressing \cite{Kvinikhidze:2001xb}.

\begin{table}[b]
 \begin{center}
 \begin{tabular}{lllllll} \hline \hline
   Set& & I & II & III & IV & V \\ \hline
   $w_{1^+}$ & [GeV$^2$] & 0.4 & 0.6 & 0.8 & 1.0 & 1.2 \\ \hline
   $w_{0^+}$ & [GeV$^2$] & 0.21 & 0.27 & 0.32 & 0.37 & 0.41 \\
   $m_{1^+}$ & [GeV]     & 1.02 & 1.01 & 1.01 & 1.00 & 0.99 \\
   $m_{0^+}$ & [GeV]     & 0.89 & 0.90 & 0.91 & 0.92 & 0.94 \\ \hline \hline
 \end{tabular}
 \end{center}
 \caption{Five parameter sets which describe core masses of nucleon
and delta, $M_{n,\rm c}=1.2$ GeV and $M_{\Delta,\rm c}=1.4$ GeV.}
 \label{parsets2}
\end{table}
 
\begin{table}[t]
 \begin{center}
 \begin{tabular}{lllrrrrr} \hline \hline
  & & $M_n$ [GeV] & Set& I & II & III & IV  \\ \hline
  $\mu_p$& [n.m.] & 0.94& & 3.05    & 2.94    & 2.86    & 2.79    \\
           &      & 1.2 &&  2.88 & 2.80 & 2.73 & 2.69 \\
  $\mu_n$& [n.m.] & 0.94& & $-$1.78 & $-$1.65 & $-$1.55 & $-$1.47  \\
            &       & 1.2 && $-$1.79 & $-$1.69 & $-$1.61 & $-$1.56 \\
 \hline \hline
 \end{tabular}
 \caption{Nucleon magnetic moments, compared betweens sets with  
  physical nucleon mass and core mass. For each pair of these sets,
  $w_{1^+}$ is equal and $w_{0^+}$ differs by at most 5 \%.}
 \label{mu_core}
 \end{center}
\end{table} 

\begin{figure}
 \begin{center}
  \epsfig{file=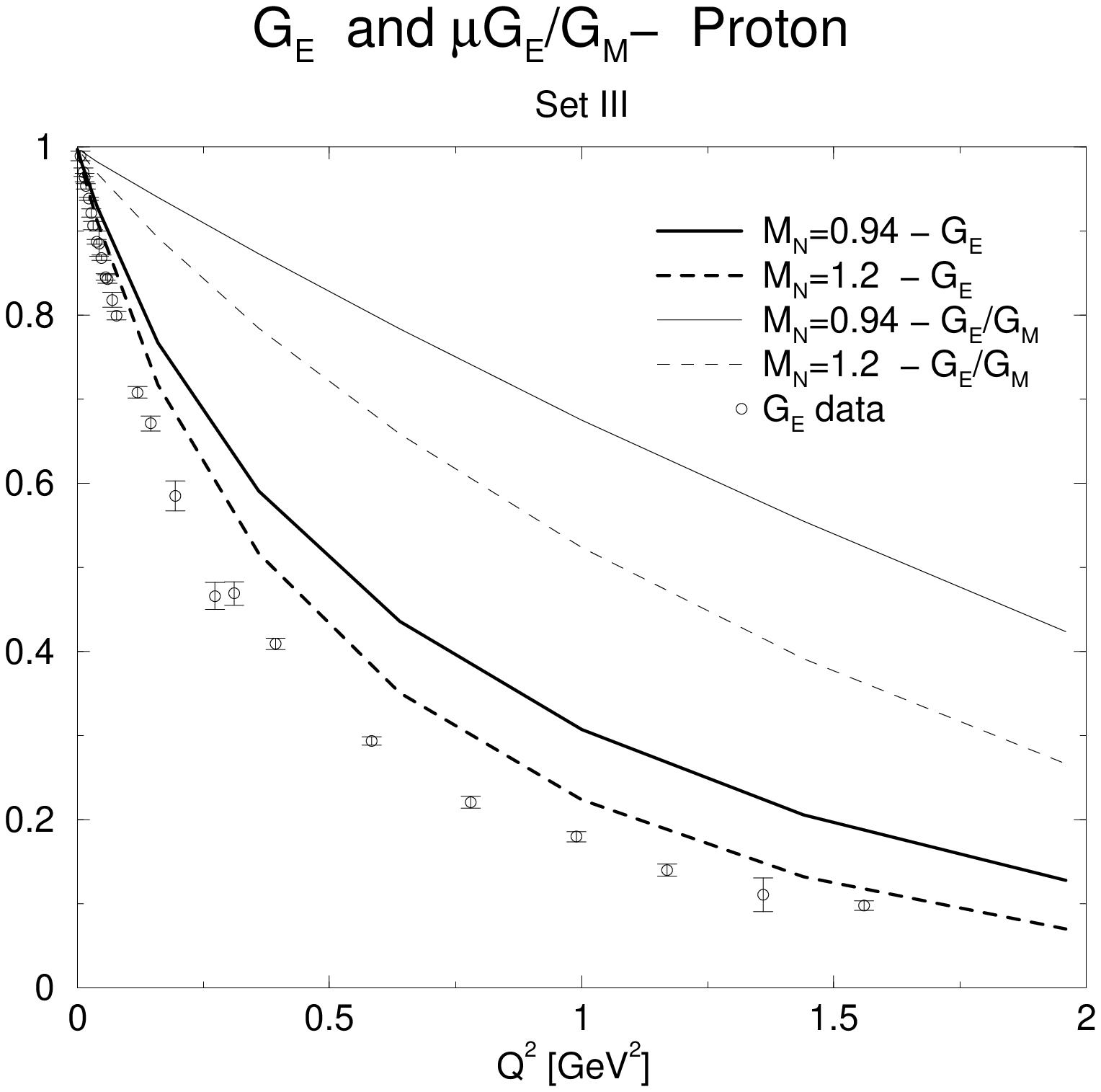,width=6cm} \hskip 2cm
  \epsfig{file=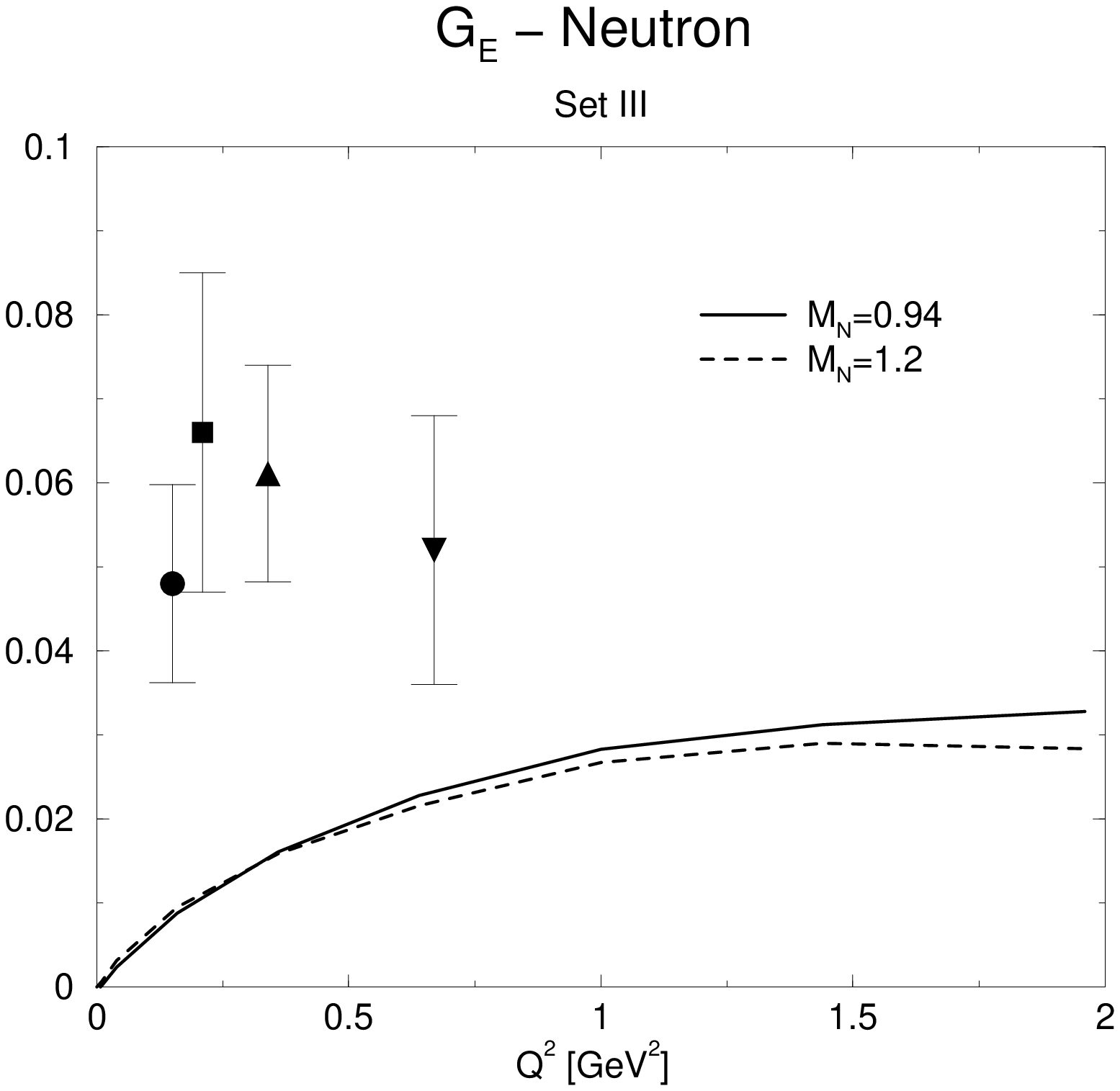,width=6cm}
  \caption{Form factor results compared for data sets describing nucleons
  with physical mass (full line)  and core mass 1.2 Gev (dashed line).
  Left panel: proton's $G_E$ and $\mu G_E/G_M$. Right panel: neutron's $G_E$.
  }
  \label{set3-fig}
 \end{center}
\end{figure}

The proton magnetic moments are slightly reduced compared to the results 
for nucleons with physical mass whereas the neutron magnetic moments
come out larger by a few percent, see Tab.~\ref{mu_core}. We present
exemplary results for the form factors $G_E$ and proton's
$G_E/G_M$ in Fig.~\ref{set3-fig}, comparing Set III from Tabs.~\ref{parsets}
and \ref{parsets2}. The characteristics found here apply to the other
data sets as well. The proton electric form factor becomes steeper thereby
approaching the experimental curve and at the same time the ratio $G_E/G_M$ 
moves even further away from the data. Since for higher core mass the
quark--photon vertex tests the
running mass function deeper in the timelike domain, its steeper derivative 
there leads to this effect.
 The neutron electric form factor
remains unchanged.

\section{Resonance contribution to the quark--photon vertex}
\label{vert_res}

In this section we describe shortly the procedure to fix a $\rho-\omega$
resonance term in the quark--photon vertex. A somewhat longer discussion
of the subject and the techniques used herein can be found in 
ref.~\cite{Roberts:2000aa}.

The transverse part of the quark--photon vertex, $\Gamma^\mu_{q,{\rm T}}$ will
certainly receive resonance contributions from the $\rho-\omega$ mesons.
Assuming isospin symmetry and neglecting the decay width, the
full vertex $\Gamma^\mu_q$ is dominated near the resonance by
the term
\bea
  \label{onsv}
  \Gamma^\mu_q(k,p) &=& f_\rho m_\rho \frac{\phi^\mu(q;Q)}{Q^2+m_\rho^2} \\
           Q &=& k-p \qquad \qquad q=(k+p)/2 \; .
\eea
Here, $m_\rho=0.77$ GeV and $f_\rho=0.216$ GeV are the mass and electromagnetic
decay constant of the $\rho$ meson. The BS amplitude ${\phi^\mu(q;Q)}$
of the $\rho$ meson is transversal ($Q\cdot \phi=0$) and obeys the canonical
normalization condition
\bea
 \label{rho_norm}
 2 Q^\mu = \frac{1}{3}\;3\;{\rm Tr} \fourint{q} \bar\phi^\mu(q;Q)\,
  S(-Q/2+q)\,\phi^\mu(q;Q) \dpart{S(Q/2+q)}{q^\mu} \; ,
\eea
if we assume that an interaction kernel for the corresponding BS equation
is independent of $Q$ (as e.g. a dressed gluon exchange between the quarks). 
The factor $1/3$ comes from the sum over the three $\rho$ polarizations,
and the factor 3 is the result of the combined flavor and color trace.
In color space, both $\rho$ and $\omega$ amplitudes are the unit matrix
$\delta_{AB}$, in flavor space we have $(\tau^3)_{ab}/\sqrt{2}$ ($\rho$)
and $(\tau^0)_{ab}/\sqrt{2}$ ($\omega$).

Away from the resonance mass--shell the corresponding contribution to the 
quark--photon vertex is not fixed uniquely. The most thorough study
of it is ref.~\cite{Maris:2000bh}, which calculates the (gluon--ladder) 
dressed quark--photon vertex for the evaluation of the pion's form factor.
The findings of ref.~\cite{Maris:2000bh} may be neatly summarized by
the following points:
\begin{itemize}
 \item $\phi^\mu(q;Q) \approx i\gamma^\mu_{\rm T}\; V_1(q^2)+
   2q^\mu_{\rm T}\;V_5(q^2)$ represents
   a good approximation to the BS solution for the $\rho$ meson
   ( $v^\mu_{\rm T}=v^\mu-Q^\mu v\cdot Q/Q^2$). It reproduces
   the mass and decay width within 5 \%.
 \item The Dirac structure $i\gamma^\mu_{\rm T}$ accounts also for the bulk
   of the resonance contribution to the pion form factor, whereas 
   terms $\sim q^\mu_{\rm T}$ provide corrections to these contributions
   on the level of 10 \%. Using the $\rho$ BS amplitude
   off its mass shell in the manner of eq.~(\ref{onsv}) gave a good 
   approximation to the quark--photon vertex resonance contributions.
 \item The Dirac structure $(q^\mu_{\rm T}-\gamma^\mu_{\rm T}
   \Slash{q}_{\rm T})\Slash{Q}$ becomes more important for intermediate
   $Q^2$ in the pion form factor but the off shell extrapolation
   of the corresponding BS amplitude structure proved to be difficult.
   Thus we neglect this term.  
\end{itemize}
We therefore adopt an 
off-shell parametrization of the resonance term in the quark--photon vertex,
\bea
 \label{resv}
 \Gamma^\mu_{q,{\rm T}}&=& \phi^\mu(q)\;
  \frac{m_\rho}{f_\rho} \frac{Q^2}{Q^2+m_\rho^2}\;e^{-\alpha\left(
  1+\frac{Q^2}{m_\rho^2}\right)} \; .
\eea
Near the $\rho$ mass shell, $Q^2=-m_\rho^2$, eq.~(\ref{resv}) reduces
to eq.~(\ref{onsv}). The exponential ensures that for high spacelike
$Q^2$ the resonance term vanishes and the quark--photon vertex
reduces to the Ball--Chiu vertex, eq.~(\ref{qvertex}). 
The resonance term vanishes also for $Q=0$ as it should be, since 
at this kinematical point the quark--photon vertex is completely fixed by
the (differential) Ward identity. 

We model the $\rho$ BS amplitude
close to the results of refs.~\cite{Maris:2000bh,Maris:1999nt}
by employing the one--parameter form
\bea
 \phi^\mu = \left(i\gamma^\mu_{\rm T} - 1.69 \frac{q^\mu_{\rm T}}{\omega_\rho}
   \right)\; \frac{{\cal F}^2(q^2/\omega^2_\rho)}{N_\rho}
\eea
The normalization constant $N_\rho$ is implicitly given by
eq.~(\ref{rho_norm}) and
the width parameter $\omega_\rho$ will
be fixed by the experimental value for the decay constant, whose
theoretical expression is
\bea
 f_\rho = \frac{1}{m_\rho}\;{\rm Tr}\;\left.\fourint{q} (-i\gamma^\mu)\,
   S (-Q/2+q)\,\phi^\mu(q)\,S(Q/2+q)\;\right|_{Q=(\vec{0},im_\rho)}\; .
\eea

The only unknown parameter which remains in eq.~(\ref{resv}) is the
constant $\alpha$ which describes the damping of the off--shell
resonance contribution. We fit it to the pion form factor in the range
$Q^2=[0,1.6]$ GeV$^2$ where the experimental data are well described
by the monopole fit \cite{Volmer:2001ek}
\bea
 F_\pi(Q^2) = \frac{1}{1+\frac{Q^2}{0.529\; {\rm GeV}^2}} \;.
\eea 
In impulse approximation, the  pion form factor is given by
\bea
 F_\pi(Q^2) &=& \frac{3}{P^2} \;{\rm Tr}\;
      \fourint{k} \;\bar \phi_\pi(k_f^2)\,S(k_1)\,\phi_\pi(k_i^2)\,
        S(k_+)\, \Gamma^\mu_q(k_+,k_-)\,S(k_-) 
    \quad\quad \\
   k_+ &=& P/2+k+Q/2\;, \qquad k_f=k+Q/4 \; , \nonumber \\
   k_- &=& P/2+k-Q/2\;, \qquad k_i=k-Q/4 \; , \nonumber \\
   k_1 &=& -P/2+k \;,   \nonumber \\
  Q&=&(0,0,|Q|,0)\;, \qquad  P=(\vec{0},i\sqrt{m_\pi^2+Q^2/4}) \;. \nonumber 
\eea
For the region of momentum transfer in consideration, the truncation
of the pion's BS amplitude to the leading amplitude which is determined 
by chiral symmetry,
\bea
 \phi_\pi(p^2) =\frac{B(p^2)}{N_\pi} \; \gamma_5 \;, 
\eea
is an excellent approximation. Indeed, the quark propagator used herein 
has been fitted 
to just give $N_\pi=f_\pi=93$ MeV, as expected physically\footnote{We 
normalize the pion BS amplitude in color space by $\delta_{AB}$ and
in flavor space by $(\tau^k)_{ab}$}.

\begin{figure}
 \begin{center}
 \epsfig{file=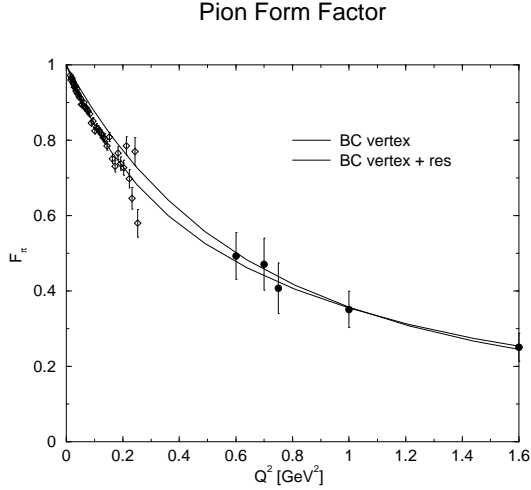,width=7cm}
 \caption{The pion form factor, calculated with the Ball--Chiu vertex 
 (thin line) and the vertex including the vector meson resonance 
 (thick line). Experimental data are from ref.~\cite{Amendolia:1986wj} 
 (diamonds) and ref.~\cite{Volmer:2001ek} (filled circles). } 
 \label{piff_f}
\end{center}
\end{figure}

As a result, we obtain $\rho$ BS amplitude width of $w_\rho^2=0.35$
GeV$^2$ which in turn gives $f_\rho=0.214$ GeV. The damping factor 
$\alpha=0.652$ results in a pion form factor as shown in Fig.~\ref{piff_f},
compared to the result obtained with only the Ball--Chiu vertex.
We remark that the latter yields a pion charge radius of 
$r_\pi^2=0.31$ fm$^2$, whereas with the resonance contribution
we obtain $r_\pi^2=0.44$ fm$^2$, in acordance with the experimental value.
Thus, about 30 \% of the charge radius is attributed to the vector mesons.
This contribution is only half of the value obtained in 
ref.~\cite{Maris:2000bh} and thus reflects the model dependence of the
off--shell extrapolation: The parametrization for the quark propagator
used here has a steeper $Z(p^2)=1/A(p^2)$ than the corresponding
renormalization function of the quark propagator obtained in
ref.~\cite{Maris:2000bh}. The $Q^2-$variation of the scalar function
multiplying the dominant Dirac structure 
$\sim \gamma^\mu_{\rm T}$ in the quark--photon vertex is therefore
already rather steep for the BC vertex used in this work and the
resonance part is seen too influence the form factor only up to
$Q^2=1$ GeV$^2$.

\section{Singularity structure of the form factor diagrams}
\label{tech}

\end{appendix}

Singularities in the diagrams are present through the
quark--photon vertex $\Gamma^\mu_q$, see eq.~(\ref{qvertex}). It contains
the scalar functions $A$ and $B$ which are defined by
\bea
 A(p^2) &=& \frac{1}{\sigma_V(p^2)}\frac{1}{p^2+M^2(p^2)} \; , \\
 B(p^2) &=& \frac{M(p^2)}{\sigma_V(p^2)}\frac{1}{p^2+M^2(p^2)}  \; , 
  \qquad (M(p^2)=\sigma_S(p^2)/\sigma_V(p^2)) \; .
\eea

We see that these functions have poles whenever $\sigma_V$ or
$p^2+M^2$ have zeros. 
The poles in $1/\sigma_V$ do not matter since in
the current matrix element diagrams, $\Gamma^\mu_q$ always appears
with quark legs, $S\Gamma^\mu_q S$, and these legs cancel the pole.
Such a mechanism is not present for the poles in
$1/(p^2+M^2)$. A numerical search revealed the following poles, being
closest to the origin in the complex $p^2$ plane:

\begin{center}
\begin{tabular}{cc} \hline \hline
  Re $p^2$ & Im $p^2$  \\
  \multicolumn{2}{c}{GeV$^2$} \\
 $-$0.067 & $\pm$ 0.207 \\
 $-$0.167 & $\pm$ 1.116 \\
 $-$0.224 & $\pm$ 1.360 \\
 $-$0.293 & $\pm$ 0.742  \\ \hline \hline
\end{tabular}
\end{center}

We see that the pole locations appear in complex conjugate pairs.
It is only the first pair of poles in the list which will have an impact
on our calculations.
To see that, let us consider the single form factor 
diagrams.

We calculate the form factors in the standard Breit frame where
\bea
 P&=&(0,0,0,i\sqrt{M^2+Q^2/4}) \nonumber \\
 P_i&=&(0,0,-|Q|/2,i\sqrt{M^2+Q^2/4}) \\
 P_f&=&(0,0,+|Q|/2,i\sqrt{M^2+Q^2/4}) \nonumber 
\eea
We start with the current matrix elements of the impulse approximation 
(diagrams (a) in Fig.~\ref{ff_f}). The quark diagram (where
the diquark is spectator) is given by
\bea
\langle J^\mu_q \rangle &=& \fourint{k} \bar \Psi (p_f,P_f)
  \begin{pmatrix} D^{-1} & 0 \\ 0 & (D^{\mu\nu})^{-1} \end{pmatrix}(k_d)
  \Gamma^\mu_q(k_q,p_q) \Psi (p_i,P_i) \\
     p_i &=& k-(1-\eta) Q/2\; , \qquad k_q=\eta P +k+Q/2  \; ,\nonumber\\
     p_f &=& k+(1-\eta) Q/2 \; , \qquad p_q=\eta P +k-Q/2  \; , \nonumber\\
     k_d &=& (1-\eta)P-k \; .\nonumber
\eea
The loop momentum $k$ is real, but the quark momenta $k_q$ and $p_q$ at the
vertex are not. Since their imaginary part, $\eta P$ grows 
with increasing $Q^2$,
the integration domain will cross the pole locations $p_{\rm pole}^2$.
The limit for $Q^2$ such that the integration domain will be free of these
poles is
\bea
 \label{Qlimit}
 Q^2 <  2\,\frac{ |p_{\rm pole}^2| - {\mbox Re}\;p_{\rm pole}^2}{\eta^2}- 
 4M_n^2 \; .
\eea
Beyond this limit, the integration path in the variable $k_4$ has 
to circumvent the poles or, alternatively, the sum of a principal value integral
with the original path and a closed contour integral encircling
the singularities must be calculated. This fact has been overlooked
in ref.~\cite{Bloch:1999ke}. While this procedure has been carried out
for {\em real} poles in ref.~\cite{Oettel:2000gc}, for {\em complex} poles
the knowledge of the wave function at {\em complex} relative momenta
between quark and diquark is required. To obtain the wave function at 
these points is in principle possible, but the implementation in a
current matrix element code is at present not feasible. 

Choosing small momentum partitioning parameters $\eta$ shifts the limit
(\ref{Qlimit}) to larger values. Due to the presence of the diquark pole,
solutions of the Faddeev equation are restricted by $\eta>1-m_{0^+}/M_n$,
in practice we have to restrict ourselves to $\eta=0.32$ to have
the Chebyshev expansion of the wave function converge for {\em both}
the Faddeev solution and the calculation of the current matrix
elements. This yields the limit $Q^2<2$ GeV$^2$, by virtue of 
eq.~(\ref{Qlimit}).

In the diquark diagram (with the quark being spectator), singularities
occur only where the photon resolves the quark loop, i.e. in the
integrals
\bea
\langle J^\mu_{0^+} \rangle &=& \fourint{k} \bar \Psi^5 (p_f,P_f)
  S^{-1}(k_q)
  \Gamma^\mu_{0^+}(k_d,p_d) \Psi^5 (p_i,P_i) \\
     p_i &=& k+\eta Q/2\;, \qquad k_d=(1-\eta) P -k+Q/2 \;,  \nonumber\\
     p_f &=& k-\eta Q/2 \;, \qquad p_d=(1-\eta) P -k-Q/2 \;,  \nonumber\\
     k_q &=& \eta P+k \nonumber \\
 \Gamma^\mu_{0^+}(k_d,p_d) &=& {\rm Tr}\,\fourint{q} \bar \chi^5({\Sc q+Q/4})
  S(q_2)\Gamma^\mu_q(q_2,q_1)S(q_1) \chi^5({\Sc q-Q/4}) S^T(q_3)\nonumber\\ \\
   q_1 &=& (p_d+k_d)/4 +q-Q/2\;, \qquad q_2 = (p_d+k_d)/4 +q+Q/2 \;, \nonumber\\ 
   q_3 &=& (p_d+k_d)/4-q \;.\nonumber
\eea
While these equations describe the contribution of the scalar diquark,
similar expressions hold for the axialvector diquark and the 
scalar--axialvector transitions. The imaginary part of the quark momenta 
$q_1$ and $q_2$ which enter the quark--photon vertex is given by
$(1-\eta)P/2$, therefore we can apply eq.~(\ref{Qlimit}) for a pole--free
integration domain upon the replacement $\eta \to (1-\eta)/2$. If
we want to calculate pole--free up to $Q^2=2$ GeV$^2$, we find
$\eta>0.36$. Thus the Faddeev solutions and these diagrams have to be 
calculated with a {\em different} momentum partitioning parameter
than the Faddeev solutions for the quark diagram. This is of course
possible since the Faddeev solutions have been obtained fully
covariantly, i.e. the full dependence of the wave function on
$p^2$ and $p\cdot P$ has been retained ($p$ is the relative quark--diquark
momentum and $P$ is the total nucleon momentum).

Singularities in the exchange kernel contributions (diagrams
(b) in Fig.~\ref{ff_f}) are present in the diagram where the photon
couples to the exchange quark. The corresponding current
matrix element is given by 
\bea
 \langle J^\mu_{ex} \rangle &=& \fourint{k}\fourint{p}
  \bar \Psi^a(k,P_f) \chi^a(p_s)
  \left(S(q_1)\Gamma^\mu_q(q_1,q_2)S(q_2)\right)^T \times\nonumber \\
 & & \qquad\qquad \bar\chi^b(k_s) \Psi^b(p,P_i)  \qquad (a,b=\{5,\mu\}) \\
 q_1 &=& -p-k+(1-2\eta)P-\frac{Q}{2}\; ,\quad p_s = k+\frac{p}{2}-(1-3\eta)\frac{P}{2}-
         (\eta-1)\frac{Q}{4} \nonumber    \\  
 q_2 &=& -p-k+(1-2\eta)P+\frac{Q}{2}\; ,\quad k_s = p+\frac{k}{2}-(1-3\eta)\frac{P}{2}+
         (\eta-1)\frac{Q}{4} \nonumber      
\eea
Here the imaginary part of the momenta $q_1$ and $q_2$ (appearing in
the quark--photon vertex) is given by $(1-2\eta)P$. Using eq.~(\ref{Qlimit})
again (with $\eta \to (1-2\eta)$), we find the condition
$\eta>0.34$ if we want to calculate this diagram without encountering
poles up to $Q^2=2$ GeV$^2$.

\end{document}